\documentclass[preprint,12pt]{elsarticle}

\usepackage{amssymb}
\usepackage{amsmath}

\usepackage{lineno}
\usepackage{amssymb}
\usepackage{amsmath}
\usepackage{mathtools}
\usepackage{amsthm}
\usepackage{multirow}
\usepackage{diagbox}
\usepackage{microtype}
\usepackage{graphicx}
\usepackage{subfigure}
\usepackage{booktabs}
\usepackage{hyperref}

\journal{Artificial Intelligence in Medicine}

\begin{document}

\begin{frontmatter}



\title{Towards a Better Model with Dual Transformer for Drug Response Prediction} 




\author[label1]{Kun Li}
\author[label2]{Jia Wu}
\author[label1]{Bo Du}
\author[label3]{Sergey V. Petoukhov}
\author[label4]{Huiting Xu\corref{cor1}}
\author[label5]{Zheman Xiao\corref{cor1}}
\author[label6,label1]{Wenbin Hu\corref{cor1}}
\cortext[cor1]{Corresponding author. hwb@whu.edu.cn, zmxiao@whu.edu.cn, annexu333@126.com}

\affiliation[label1]{organization={School of Computer Science},
            addressline={Wuhan University}, 
            city={WuHan},
            postcode={430037}, 
            state={Hubei},
            country={China}}
            
\affiliation[label2]{organization={School of Computing},
            addressline={Macquarie University}, 
            city={Sydney},
            postcode={201101}, 
            state={NSW},
            country={Australia}}  

\affiliation[label3]{organization={Department of Biomechanics},
            addressline={Mechanical Engineering Research Institute of the Russian Academy of Sciences}, 
            city={Moscow},
            postcode={101990},
            country={Russia}}

\affiliation[label4]{organization={Department of abdominal Oncology},
            addressline={Hubei Cancer Hospital}, 
            city={WuHan},
            postcode={430037}, 
            state={Hubei},
            country={China}}

\affiliation[label5]{organization={Department of Neurology},
            addressline={Renmin Hospital of Wuhan University}, 
            city={WuHan},
            postcode={430037}, 
            state={Hubei},
            country={China}}
            
\affiliation[label6]{organization={Wuhan University Shenzhen Research Institute},
            city={Shenzhen},
            postcode={518000}, 
            state={Guangdong},
            country={China}}
            
\begin{abstract}
At present, the mainstream methods of drug response prediction (DRP) tasks are based on convolutional neural networks, graph neural networks, etc. These approaches focus only on the structural information of drug molecules, while ignoring to some extent the pharmacological and chemical properties represented by edges and nodes.  In addition, fragment-based molecular encoding methods and multi-omics cell encoding methods face challenges due to inconsistencies in practical predictions. This is primarily attributed to variations in the size of molecular features across datasets and difficulties in collecting multi-omics data for a single cell type, thereby impeding their applications. Moreover, in the field of cell biology, the existing methods are unable to completely perceive the genomics sequences of cell lines at once. The above problems lead to potential security risks and low prediction accuracy. To address these problems, we propose a dual transformer structure with edge embedding for the DRP task, named TransEDRP. In the drug branch, a graph transformer with the edge embedding module is proposed to extract the pharmacochemical properties and structural information of drug molecules; while in the cell lines branch, a transformer-based genomic encoder module is proposed to globally extract genomic sequences. A multi-modal soft fusion module is then designed to fuse the drug molecules and genomics to predict the half-maximal inhibitory concentration score. We conduct experiments on a wide range of datasets to show that our method achieves state-of-the-art results. Furthermore, the results of the chirality experiment validate the ability of our model to reduce drug safety risk. The codes are available at \href{https://github.com/DrugD/TransEDRP}{https://github.com/DrugD/TransEDRP}.

\end{abstract}

\begin{graphicalabstract}
\includegraphics[width=1\textwidth]
{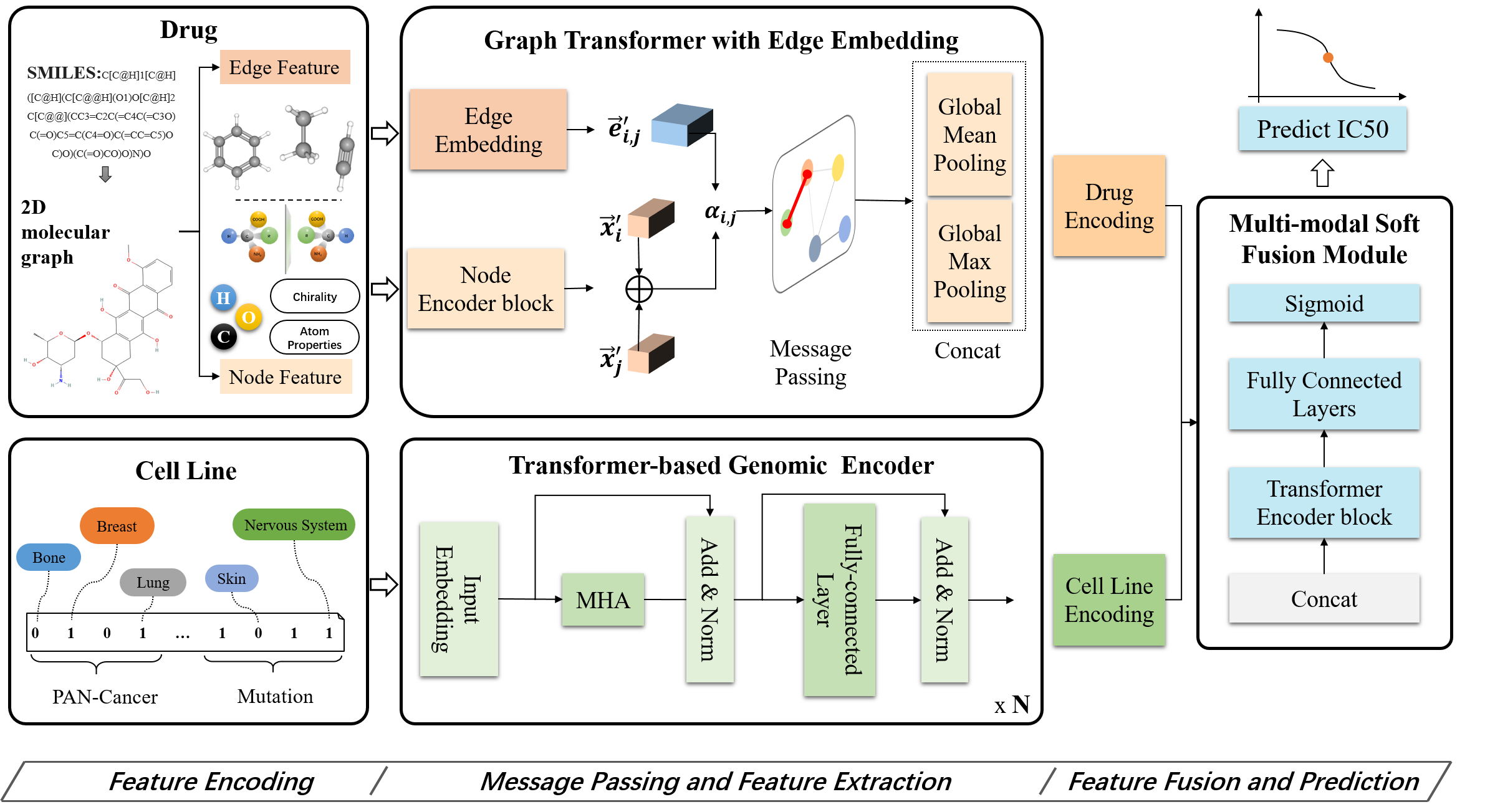}
\end{graphicalabstract}

\begin{highlights}
\item In drug discovery, drug response prediction (DRP) plays a crucial role.

\item We propose a dual transformer model with edge embedding for DRP.

\item It leverages the chemical bond of drugs and their pharmacological properties.

\item Experimental results on five datasets demonstrate the state-of-the-art (SOTA) performance achieved by TransEDRP.

\item Chiral experiments show that this model has the potential to improve the safety of drug discovery.

\item By considering the performance, safety, stability, and feasibility in drug screening, a comparison of existing SOTA methods indicates that TransEDRP is a better model.

\end{highlights}

\begin{keyword}
Drug Respond Prediction \sep Graph Transformer \sep Chemical Bond \sep Chiral Isomer
\end{keyword}

\end{frontmatter}



\section{Introduction}
\label{Introduction}

\begin{figure}[ht!]
    \centering
    \includegraphics[width=0.98\linewidth]{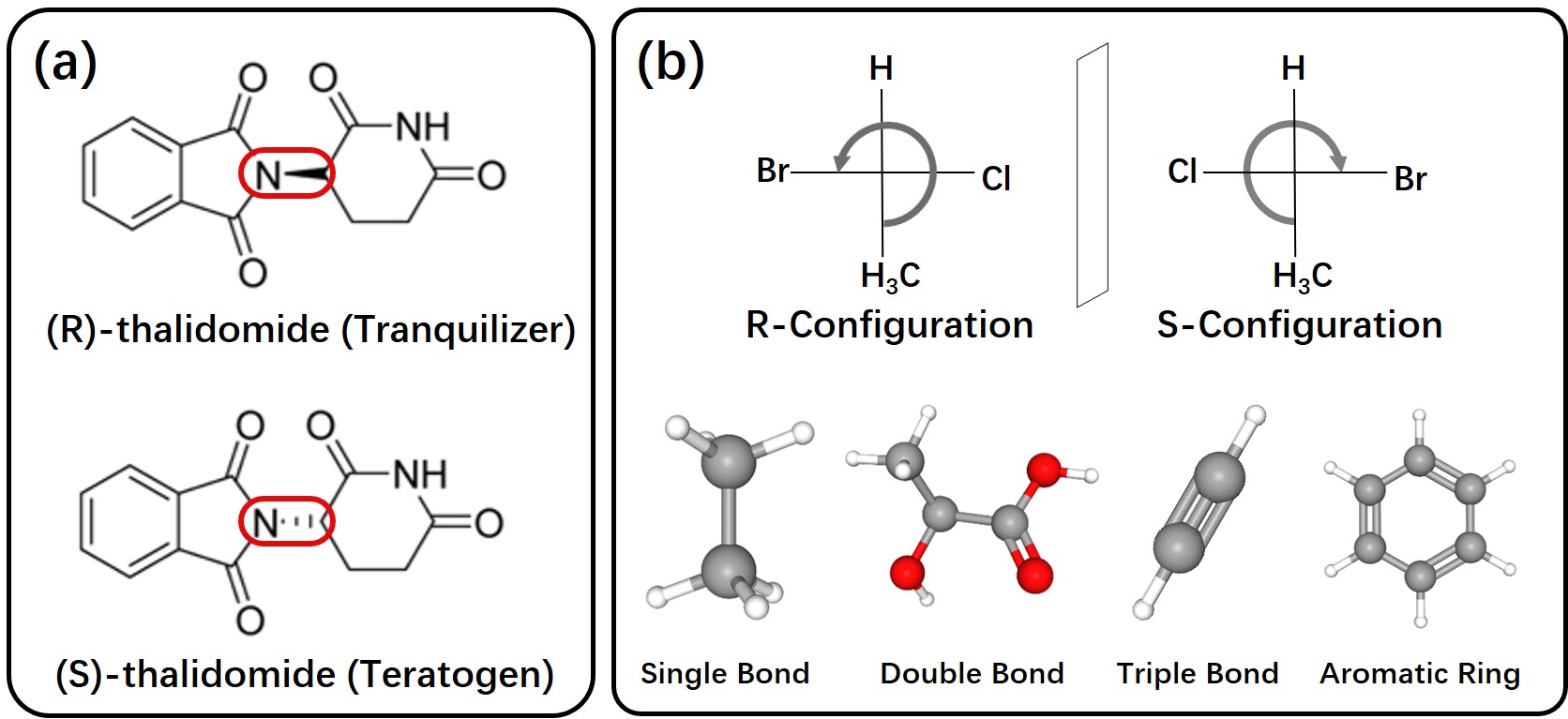}
    \caption{Fig. (a) shows a pair of chiral isomers as an example. Fig. (b) shows the edge features including covalent bonds, aromatic rings, etc., along with the chirality of the atoms.}
   	\label{fig:intro1}
\end{figure}

The task of drug response prediction (DRP) has been studied by many scientists, as it is one of the aspects of preclinical studies in drug screening \cite{drugdevelopmentplanning}. Compared with methods based on non-graph computational deep learning, graph-based methods have achieved better performance on the task of drug molecule representation \citep{NN2, threemolrep, HGRL-DTA, 33}. For example, DeepCDR \citep{DeepCDR}, as a hybrid graph convolutional network, integrated multi-omics profiles and explored the intrinsic chemical structures of drugs for DRP. In particular, GraTransDRP \citep{GraTransDRP}, which combines graph and transformer methods for DRP, has achieved state-of-the-art (SOTA) results to date. Notably, the existing methods above focus only on the structural information of drug molecules while ignoring the pharmacological and chemical properties (covalent bonds, chiral isomers, rings, etc.) to some extent.  For example, Fig. \ref{fig:intro1} (a) shows two chiral isomers of Thalidomide. Among them, the difference between the R/S configuration is the chiral feature of one carbon atom (marked by a red circle, see \ref{sec:chiralConcepts} for details). Existing methods ignore the difference in the chiral features of the atoms, meaning that chiral isomers of the drug have the same graph encoding; thus these two different molecules have the same response prediction value.

In addition, there are many types of atoms and chemical bonds in a molecule: as shown in Fig. 
\ref{fig:intro1} (b), there are single bonds, double bonds, etc. 
However, existing methods ignore the effect of different chemical bonds on the drug, meaning that different molecules may be encoded in the same graph structure. It should be noted here that these molecules with similar graph structures but different chemical bonds will most likely have quite different efficacy for the same cell lines. Therefore, ignoring the feature differences in chemical bonds could reduce the accuracy of models. In addition to the cases mentioned above, there are numerous examples of a seemingly small structural modification having a huge impact on the effect \citep{activitycliff}. Therefore, pharmacological and chemical properties such as chirality, covalent bonds, and aromatic rings are taken into consideration, which can aid in improving chirality security and prediction accuracy.

\begin{figure}[ht!]
    \centering
	\includegraphics[width=0.98\linewidth]{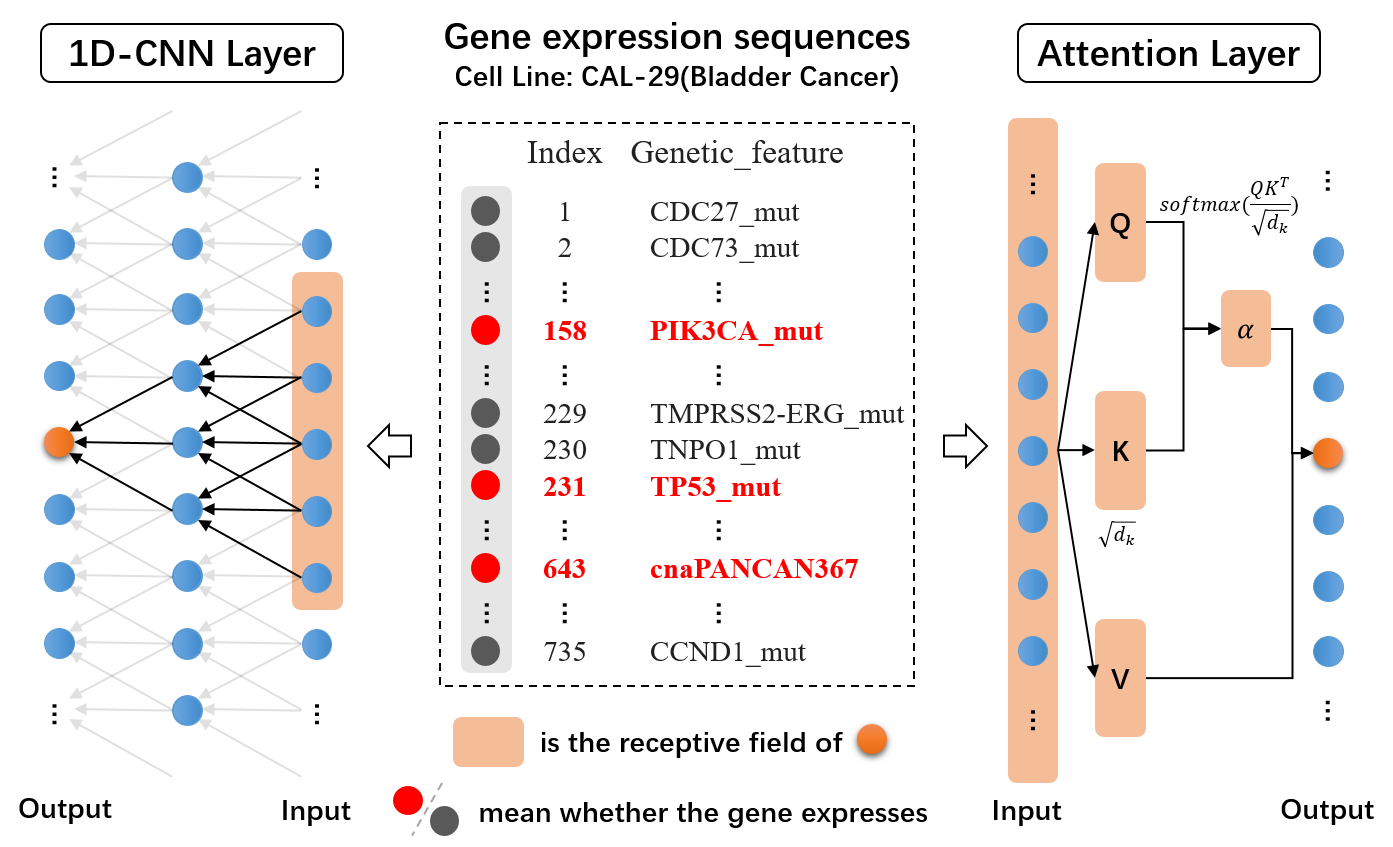}
	\caption{The gene expression sequence of CAL-29, which is derived from transitional cell carcinoma of the bladder. The perceived fields on the sequence using 1D-CNN and the Attention layer are also compared.}
	\label{fig:intro2}
\end{figure}


The other input of the DRP task is the genomic sequences of cell lines. The existing methods based on 1D-CNN \citep{tcnns,110bGCNforDRP,DeepCDR} achieve good performance, but ignore two important features of the gene sequence in cell biology, which are position independence (the position of the gene fragments does not carry information) and global complex relationships (it is vital to know which of the fragments in the overall sequence are mutated together).  These two features are important for the design of gene-sequence feature extraction methods. For example, Fig. \ref{fig:intro2} illustrates the genomic sequence extraction process for a cell line named CAL-29. Red text indicates that the gene fragment is mutated during expression, and gray is normal.  In the 1D-CNN layer, the actual receptive field of a node in the output layer contains only five gene fragments in the input layer. Therefore, the 1D-CNN cannot globally perceive all mutated gene fragments and effectively mine the interrelationships between sequences, which can lead to the extraction of biased features. So, is there a method that can perceive all information of the gene sequence at once? It is well-known that the attention mechanism is widely used in the fields of natural language processing and image analysis, as it allows each element of the input to participate in the computation of all other elements, which enables it to effectively fit the global relationships of gene sequences. Could we introduce the attention mechanism into genomics sequence extraction methods to improve the ability of global relationship modeling?

In this paper, we attempt to address the above problems by proposing a dual-branch structure \textbf{Trans}former model with \textbf{E}dge Embedded for \textbf{D}rug \textbf{R}esponse \textbf{P}rediction, called TransEDRP. We first propose a graph transformer with an edge embedding module, which combines the features of edges and nodes, and can consider the pharmacological impact of chirality and chemical bonds on drug molecules. To the best of our knowledge, this represents the first time that these features have been encoded in the graph structure representation of drug molecules. We further propose a transformer-based genomic encoder module to solve the problem of existing methods being unable to perceive the genomics sequences of all cell lines simultaneously. As shown in Fig. \ref{fig:intro2}, the attention mechanism allows each element to participate in the representation of all other elements, so that important features can be extracted from the whole sequence. 

Extensive overall experiments with many existing methods are conducted on almost all relevant publicly available datasets, comprising a total of over 500,000 real data points. These results show that TransEDRP achieved a 22.67\% improvement in the average root mean square error (RMSE) metric across all datasets compared to the baseline method tCNNs \citep{tcnns} and also achieved the SOTA level. Furthermore, the chirality experiment shows that our model can effectively distinguish the efficacy of chiral isomers on cell lines. By considering the performance, safety, stability, and feasibility in drug screening, a comparison of existing SOTA methods indicates that TransEDRP is a better model for the DRP task. Our main contributions can be summarized as follows:


\begin{itemize}
\item We propose TransEDRP in order to take full advantage of the pharmacological and chemical properties of drug molecules, as well as to extract features of genomic sequences globally.
\item We conduct chirality experiments and extensive overall experiments on five publicly available datasets. The results demonstrate that TransEDRP can effectively improve both the accuracy and security of drug response prediction tasks with high stability and feasibility. 
\end{itemize}

\section{Related work}

With the completion of large-scale projects on drug response in cell lines, such as GDSC \citep{GDSC} and CCLE \citep{CCLE}, the study of DRP became possible. Numerous DRP methods have been proposed, which can be classified as 1DCNN-based, Graph-based, and Transformer-based methods.

The non-graph-based methods generally use CNN and MLP for information extraction, with examples including CDRScan \citep{CDRscan}, tCNNs \citep{tcnns},  MOLI \citep{MOLI}, etc.
CDRscan \citep{CDRscan}, the first deep learning method for the DRP task,  uses genomic mutation as a cancer cell profile. tCNNs \citep{tcnns} and MOLI \citep{MOLI} adopt two convolution networks to learn the representations of drug and cancer cell lines. These approaches employ simplified molecular input line entry specification (SMILES) sequences of drugs and genomic mutation data of cell lines. These methods ignore the drug's pharmacological features and the molecule's structural features. The Graph-based methods focus on converting drug molecules into graph structures and then using Graph Neural Networks (GNNs) for graph representation rather than extracting features directly from SMILES strings. DeepCDR  \citep{DeepCDR} is a hybrid graph convolutional network that integrates multi-omics profiles and was the first to explore the intrinsic chemical structures of drugs.  GraphDRP \citep{110bGCNforDRP} used GNN models \citep{13dGIN,11dGCN,GAT} achieves superior performance compared to traditional machine learning methods. TGSA \citep{TGSA} integrated STRING protein-protein association networks to build cell line graphs for exploring gene relationships.  Based on the framework established by TGSA, HMM-GDAN \citep{HMM_GDAN} integrated multi-view and multi-scale graph duplex-attention networks, resulting in promising results.



Moreover, in contrast to GNNs, the Transformer avoids introducing any structural inductive bias at intermediate layers \citep{GraphTransformerForDR, structuretransformer}, thereby addressing the expressivity limitation of GNNs \cite{AGCNforDD}. GraTransDRP \citep{GraTransDRP} combines GNNs and Transformer; this approach is primarily designed to reduce the error associated with the approximation of heterogeneous graphs to homogeneous graphs. DeepTTA \citep{DeepTTC} and SubCDR \cite{SubCDR} decompose drug molecules into fragments, encode the drug molecules using fragment probabilities, and then use the CNN and Transformer to perform drug feature extraction. Many existing methods still follow the 1DCNN-based method to perform feature extraction of drug and cell line sequences, which may reduce the accuracy of drug response prediction.



\section{Method}

\begin{figure*}[t!]
    \centering
    \includegraphics[width=0.99\textwidth]{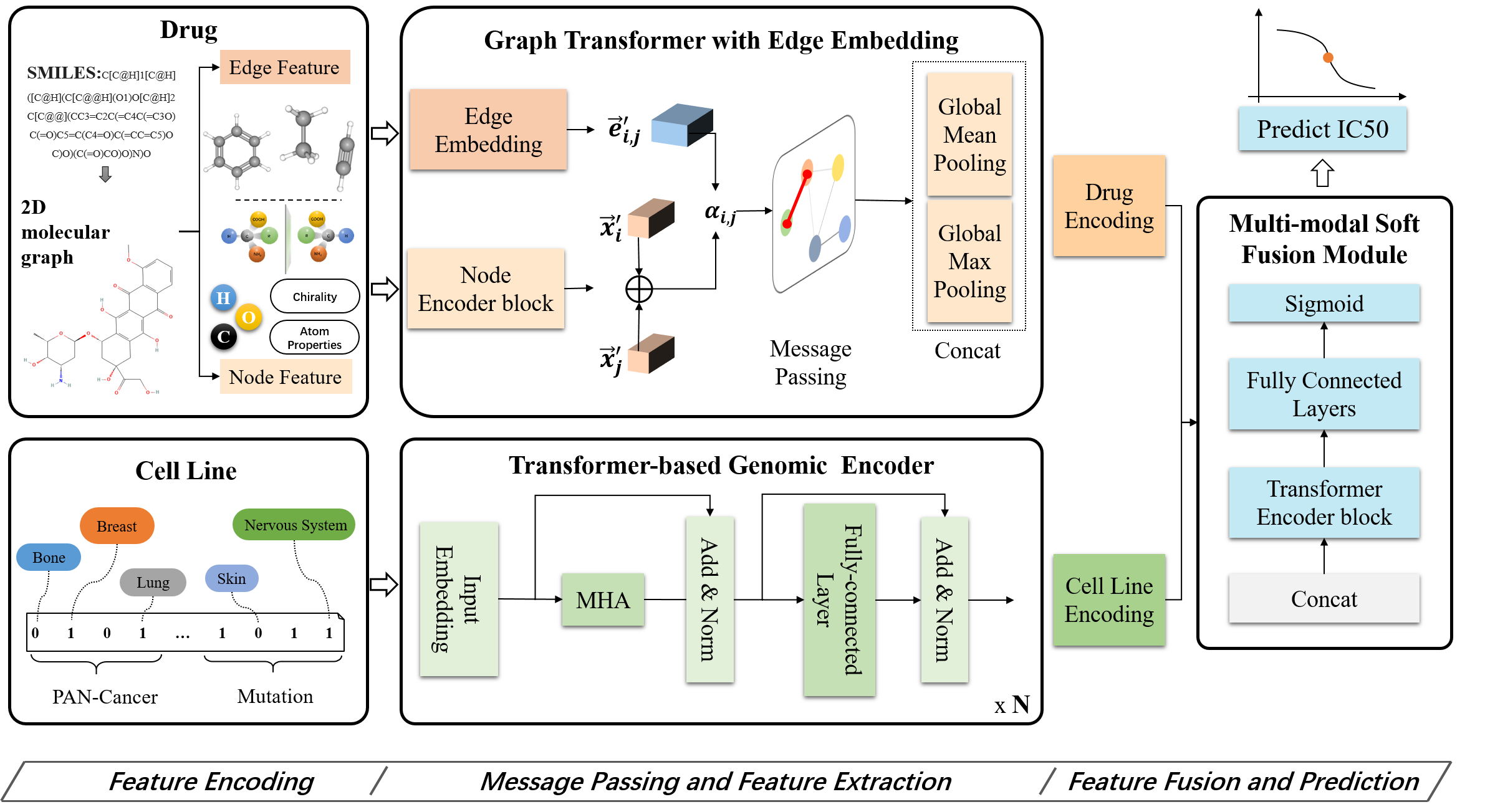}
    \caption{Illustration of the TransEDRP framework, which consists of three components: the graph transformer with edge embedding module, the transformer-based genomic encoder module, and the multi-modal soft fusion module. }
    \label{fig:model}
\end{figure*}

\subsection{Framework}

In this paper, an end-to-end deep learning model named TransEDRP is proposed to predict drug response. As shown in Fig.  \ref{fig:model},  the model comprises three modules. The graph transformer with edge embedding module is proposed to extract the higher-order semantics of the drug molecule graph,  while the transformer-based genomic encoder module is proposed to extract the genomics of the cell lines; these two types of features are then fused to predict the half maximal inhibitory concentration ($IC_{50}$) using the multi-modal soft fusion module.  The inputs of the DRP task are the drugs in graph format and the cell line sequences. As shown in Appendix Table \ref{tab:bondE}, the improvement over the mainstream encoding methods is that chemical bonds are encoded as edges in the drug molecule graph. In addition, the chirality of the atoms is added for the representation of nodes.

\subsection{Graph transformer with edge embedding module}

\label{sec:3.2}


The graph transformer with edge embedding module is made up of transformer encoder layers  and graph message passing layers. The transformer encoder layers are in turn made up of multi-head self-attention (MHA) and feed-forward network (FFN), and are used to globally extract the node features. After the node features are extracted in the transformer encoding layer, the graph neural network combines the node features and edge features for message passing on the molecular graph. 

In the graph representation of a drug molecule, the node feature matrix is denoted as $\mathbf{X}= \left \{ \vec{x} _1,\vec{x}_2,...,\vec{x}_N \right \} \in \mathbb{R}^{N \times d}$ and the edge feature matrix in one-hot format is $\mathbf{E}\in \mathbb{R}^{M \times d_e }$; here, $N,M$ represent the number of nodes and  edges, while $d,d_e$ are the feature sizes of the nodes and edges. To facilitate message passing between edge features and node features, a learned mapping  matrix $\mathbf{W}_{e} \in \mathbb{R}^{d_e \times d }$ is used to map $\mathbf{E}$  to $\mathbf{E}^\prime  = \left \{ \vec{e}^{\,\prime}_1,\vec{e}^{\,\prime}_2,...,\vec{e}^{\,\prime}_M  \right \}   \in \mathbb{R}^{M \times d } $, so that the edge features have the same dimensions as the nodes. All learned weight matrices require bias unless otherwise specified.

In the self-attention, the input matrix $\mathbf{X}$ can be transformed into query, key,  and value matrices, denoted as $\mathbf{Q},\mathbf{K},\mathbf{V} \in \mathbb{R}^{N \times d}$. The attention scores are then normalized by the softmax function, which can be defined as follows:
\begin{equation}
      \mathrm{Attention}(\mathbf{Q,K,V}) = \mathrm{softmax} \bigg(\frac{\mathbf{Q}\mathbf{K}^T}{\sqrt{d}}\bigg)\mathbf{V}
      \label{alg: attention}
\end{equation}
\noindent where $T$ denotes the transposition operation, while  $\mathbf{Q ,K ,V}$  are obtained by linear transformation of $\mathbf{X}$. In multi-head attention (MHA), the information will be represented in different subspaces, which can be defined as follows:
\begin{equation}
      \mathrm{MHA}(\mathbf{Q,K,V}) = \left [ \mathrm{head}_1,\mathrm{head}_2,...,\mathrm{head}_h \right ] \mathbf{W}^O
      \label{alg: MHA}
\end{equation}

\begin{sloppypar}
\noindent where $\mathrm{head}_{ i\in\left [1,h  \right ] } =\mathrm{Attention}\big(\mathbf{QW}^{Q}_i, \mathbf{KW}^{K}_i,\mathbf{VW}^{V}_i\big)$, $\mathbf{W}^{Q}_i$, $\mathbf{W}^{K}_i$, $\mathbf{W}^{V}_i \in \mathbb{R}^{d \times d / h }$, $\mathbf{W}^{O}_i \in \mathbb{R}^{d \times d}$ are learned matrices, and $\left [ \cdot \right ] $ denotes the feature dimension concatenation operation.
\end{sloppypar}

Moreover, $\mathcal{N}$ denotes the layer normalization, $\mathcal{D}$ denotes the dropout, and $\mathcal{F}$ denotes the feed-forward network. Therefore, the node features $\mathbf{X}^{\prime} = \left \{ \vec{x}^{\prime}_1,\vec{x}^{\prime}_2,...,\vec{x}^{\prime}_N  \right \}   \in \mathbb{R}^{N\times d}$ extracted from the transformer encoder layer $\Psi$ can be calculated as follows:
\begin{equation}
    \begin{aligned}
         \mathbf{Z}  & = \mathcal{N} (\mathbf{X}+\mathcal{D}(\mathrm{MHA}(\mathbf{X},\mathbf{X},\mathbf{X})) ) \\ \mathbf{X}^{\prime} & = \Psi\left ( \mathbf{X} \right )   = \mathcal{N} (\mathbf{Z}+\mathcal{D}(\mathcal{F} (\mathbf{Z})))
    \end{aligned}
    \label{alg: transformer}
\end{equation}
 


The graph network with edge embedding is designed to allow the edge features to constrain the aggregation weights of neighbors to facilitate local node representation. As discussed in Section \ref{sec:3.2},  the two inputs are the node features $\mathbf{X}^{\prime}$ encoded by Equation \ref{alg: transformer} and  the edge features $\mathbf{E}^{\prime}$. The attention score $\alpha_{i,j}$ between two nodes $\vec{x}^{\prime}_i$ and $\vec{x}^{\prime}_j$ with the embedding of the corresponding edge $\vec{e}^{\,\prime}_{i,j}$  can be calculated as follows:
\begin{equation}
\alpha_{i,j}=
        \frac{
        \exp\left(\mathcal{F} \left( \left [   \mathbf{W}_{1} \vec{e}^{\,\prime}_{i,j}, \vec{x}_{j}^{\prime} \oplus  \vec{x}_j^{\prime}   \right ]   \right) \right)}
        {\sum_{k \in \mathcal{M}(i) }
        \exp\left(\mathcal{F}\left(
        \left [ \mathbf{W}_{1}  \vec{e}_{i,k}^{\,\prime},    \vec{x}_{j}^{\prime} \oplus  \vec{x}_k^{\prime}  \right ] \right)\right)} 
    \label{equ:attenionEdge}
\end{equation}
\noindent where $\mathcal{M}(i)=\mathcal{N}(i)\cup \{ i \}$ represents the first-order neighbors of node $i$ (including node $i$),  $\oplus$ is the addition operation in feature dimension, and the shared linear transformation weight matrix $\mathbf{W}_{1} \in \mathbf{R}^{d \times d} $ is applied to every edge. Inspired by the node attention calculation method in GAT, the attention score $\alpha_{i,j}$ calculated by Equation \ref{equ:attenionEdge} is utilized in graph message passing, resulting in the following node feature representation:
\begin{equation}
   \vec{x}_{i}^{\prime\prime}=   \mathop{\Big|\Big|}\limits_{k=1}^{K}  \sigma\left(\sum_{j \in \mathcal{M}(i)} \alpha_{i,j}^{k} \mathbf{W}^{k} \vec{x}^{\prime}_{j}\right)
\end{equation}
\noindent where $\mathbf{W}^{k}$ is the corresponding input linear transformation’s weight matrix, $\sigma(\cdot )$ is the LeakyReLU activation function, $k \in [1,K]$ is the index of the multi-heads, and $K$ is the number of attention heads, 
$||$ denotes the feature dimension concatenation operation. The node features $\mathbf{X}^{\prime\prime} = \left \{ \vec{x}^{\prime\prime}_1,\vec{x}^{\prime\prime}_2,...,\vec{x}^{\prime\prime}_N  \right \}   \in \mathbb{R}^{N\times Kd }$ are extracted from the graph network with edge embedding.






As is well-known, a single-layer graph network is often unable to achieve  excellent results. Therefore, multiple GNN layers are required  to obtain information about high-order neighbors. Referring to Equation \ref{alg: transformer},  the node features $\mathbf{X}^{\prime\prime}_{att} = \left \{ \vec{h}_1,\vec{h}_2,...,\vec{h}_N  \right \}   \in \mathbb{R}^{N\times Kd}$ are computed using MHA before GNN.
However, it should be noted that the edge features are only encoded for first-order nodes, and that reusing the edge features for multi-order information aggregation may destroy the uniqueness of the edge constraints, leading to degradation of the model performance. Thus, a vanilla GNN without edge embedding is employed to aggregate information, which can be expressed as follows:
\begin{equation}
    \vec{h}_{i}^{\prime}=  \sigma\left(\sum_{j \in \mathcal{N}_{i}} \mathbf{W}_{2} \vec{h}_{j}\right) 
\end{equation}
\noindent where $\mathbf{W}_{2} \in \mathbb{R}^{Kd\times  Kd}$ is the linear transformation matrix, while $\mathbf{X}_{g} = \left \{ \vec{h}^{\prime}_1,\vec{h}^{\prime}_2,...,\vec{h}^{\prime}_N  \right \}   \in \mathbb{R}^{N\times Kd} $ denotes the node features. Due to the different numbers of nodes in the graph, regular graph representation vectors need to be obtained via graph pooling before graph representation. For graph pooling, the global average pooling and global maximum pooling of all nodes $\mathbf{X}_{g}$ in a graph are concatenated together for graph representation, as follows:
\begin{equation}
     \mathbf{X}_{drug} = \mathrm{softmax} \Big(\mathbf{W}_{3}^{\mathrm{T}}   \left [ \underset{max}{\mathcal{P}}  \left ( \mathbf{X}_g \right )  , \underset{mean}{\mathcal{P}} \left ( \mathbf{X}_g \right )  \right ] \Big) 
\end{equation}
\noindent where $\mathbf{W}_{3} \in \mathbb{R}^{2Kd \times 
d_{\mathrm{drug}}}$ denotes a learned matrix, while $\underset{max}{\mathcal{P}} $ and $\underset{mean}{\mathcal{P}}$ denote the graph max pooling and graph mean pooling respectively. We can then obtain the graph pooling representation $\mathbf{X}_{drug}\in \mathbb{R}^{L\times d_{\mathrm{drug}}} $; here, $L$ is the number of graphs, i.e. the number of drug molecules (each graph contains a different number of atoms, and the number of nodes in all graphs is $N$).

\subsection{Transformer-based genomic encoder module}

\begin{sloppypar}
The gene sequences are coded to obtain $ \mathbf{X}_{cell} = \left \{ \vec{c}_1,\vec{c}_2,...,\vec{c}_L \right \} \in \mathbb{R}^{L \times d_c} $, which is represented by the mutation features in a large number of genetic segments. To facilitate the perception of all gene sequences simultaneously, the overall mutational information of the cell line  $ \mathbf{X}_{cell}^{\prime} = \left \{ \vec{c}^{\,\prime}_1,\vec{c}^{\,\prime}_2,...,\vec{c}^{\,\prime}_L \right \} \in \mathbb{R}^{L \times d_c} $ is acquired with reference to Equation \ref{alg: transformer}, expressed as follows:
\end{sloppypar}

\begin{equation}
    \mathbf{X}_{cell}^{\prime}  = \sigma(\textbf{W}_{4} (\underbrace{\Psi \left ( \dots  \left ( \Psi\left ( \mathbf{X}_{cell}  \right )  \right )  \right ) }\limits_{n} ))
\end{equation}
\noindent where $n$ is the number of $\Psi(\cdot)$, ${h_i}$ denotes the number of MHA heads in the $i$-th transformer encoder layer (refer to Equation \ref{alg: MHA}), $\mathbf{W}_{4} \in \mathbb{R}^{d_c \times d_{\mathrm{cell}}}$ denotes the transformation matrix,  and   $\sigma(\cdot )$ denotes the ReLU activation.  We then obtain $\mathbf{X}_{cell}^{\prime\prime} =\left \{ \vec{c}^{\,\prime\prime}_1,\vec{c}^{\,\prime\prime}_2,...,\vec{c}^{\,\prime\prime}_L \right \} \in \mathbb{R}^{L \times d_{\mathrm{cell}}} $ as  the output of the cell lines branch before feature fusion.

\subsection{Multi-modal soft fusion module}

At this stage, we have two inputs needing to be fused, namely the molecule graphs and cell line sequences, which come from different feature spaces. A multi-modal soft fusion module made up of the transformer layer and the FFN is designed for the efficient and global aggregation of features, and is superior to the traditional methods of hard summing or concatenation. Specifically, to increase the  entanglement of the representations from different modalities, the transformer encoder layer $\Psi$ is used to unify the distribution of $\mathbf{X}_{drug}$ and $ \mathbf{X}_{cell}^{\prime\prime} $, which is beneficial for prediction. The soft fusion module can be represented as follows:
\begin{equation}
       \mathbf{Y}_{pred} =  \mathrm{sigmoid} \Big( \mathcal{F}\big(\Psi (\left  [ \mathbf{X}_{drug}, \mathbf{X}_{cell}^{\prime\prime} \right ])\big)\Big)
\end{equation}


The output $\mathbf{Y}_{pred}= \left \{ y^{\prime}_1,y^{\prime}_2, ...,y^{\prime}_L \right \} $ is the prediction of the 
$IC_{50}$ scores. We denote $\mathbf{Y}= \left \{ y_1,y_2, ...,y_L \right \}$  as the true label of $IC_{50}$ scores. Finally, for the regression constraints, we define the loss function expressions as follows:
\begin{equation}
    \mathcal{L} = \frac{\sum_{i=1}^{L}(y^{\prime}_i - y_i)^2}{L}
\end{equation}
\noindent where $\mathcal{L}$ is the result of the mean squared loss function.



\section{Experiment}

\subsection{Experimental settings}

We employed the following five pan-cancer drug response datasets (see Appendix Table \ref{tab:dataset}). To evaluate the performance of different methods on these datasets, we utilize three metrics: Pearson's correlation coefficient (Pearson), Spearman's correlation coefficient (Spearman), and root mean square error (RMSE). In addition, we choose the \textbf{Isomeric SMILES} format containing stereo, and isotopic information to support our chirality experiments (see \ref{apd:Chiralsecuritytest}).

The node encoding method follows the work of GraphDRP \citep{110bGCNforDRP}, to which we have added chirality to the drug molecular. Similarly, the edge representation shown in Appendix Table \ref{tab:bondE} is encoded in the one-hot format with the chirality.  The cell line genomic includes mutation and copy number aberration (as shown in Fig. \ref{fig:intro2}), created as the one-hot format with a vector of 735 dimensions.

\subsection{Overall  experiment}
\label{sec:4.2}

\begin{table*}[t!]
\centering
\caption{Performance comparison of existing methods on five datasets. The best performance is marked in \textbf{bold}, and the second best experimental result is indicated in \textit{italics}.}
\setlength{\tabcolsep}{10pt}
\resizebox{\textwidth}{!}{%
\begin{tabular}{ccccccccc}
\toprule[1pt]
Method & Metric & GDSCv1 & GDSCv2 & CTRPv1 & CTRPv2 & GCSI & Mean \\ \hline
  & RMSE$\downarrow$ & 0.0257 & 0.0351 & 0.0284 & 0.0677 & 0.0371 
 & 0.0388
 \\
tCNNs  & Pearson$\uparrow$ & 0.9189 & 0.8558 & 0.7165 & 0.3477 & 0.8551 & 0.7388
\\
\citep{tcnns} & Spearman$\uparrow$ & 0.9010 & 0.8240 & 0.5604 & 0.3497 & 0.8694 & 0.7009
\\ \midrule
\specialrule{0em}{1.5pt}{1.5pt}
\midrule
 & RMSE$\downarrow$ & 0.0253 & 0.0266 & 0.0303 & 0.0368 & 0.0422 & 0.0322
\\
DeepCDR & Pearson$\uparrow$ & 0.9243 & 0.9188 & 0.6670 & 0.8364 & 0.8953 & 0.8484
\\
\citep{DeepCDR} & Spearman$\uparrow$ & 0.9034 & 0.8902 & 0.5061 & 0.8518 & 0.8781 & 0.8059
\\ \hline
 & RMSE$\downarrow$ & 0.0256 & 0.0270 & 0.0288 & 0.0363 & 0.0372 & 0.0310
\\
GraphDRP(GCN) & Pearson$\uparrow$ & 0.9196 & 0.9139 & 0.7073 & 0.8398 & 0.9173 & 0.8596
\\
\citep{110bGCNforDRP} & Spearman$\uparrow$ & 0.9011 & 0.8891 & 0.5510 & 0.8601 & 0.8969 & 0.8196
\\ \hline
 & RMSE$\downarrow$ & 0.0234 & 0.0251 & 0.0288 & 0.0357 & 0.0343 & 0.0295
\\
GraphDRP(GIN) & Pearson$\uparrow$ & 0.9356 & 0.9271 & 0.7090 & 0.8461 & 0.9331 & 0.8702
\\
\citep{110bGCNforDRP} & Spearman$\uparrow$ & 0.9186 & 0.9034 & 0.5488 & 0.8664 & \textit{0.9159} & 0.8306
\\ \hline
 & RMSE$\downarrow$ & 0.0236 & 0.0256 & 0.0283 & 0.0358 & 0.0362 & 0.0299
\\
GraphDRP(GAT) & Pearson$\uparrow$ & 0.9323 & 0.9227 & 0.7171 & 0.8446 & 0.9219 & 0.8677
\\
\citep{110bGCNforDRP} & Spearman$\uparrow$ & 0.9153 & 0.8984 & 0.5694 & 0.8652 & 0.9058 & 0.8308
\\ \hline
 & RMSE$\downarrow$ & 0.0233 & 0.0252 & 0.0284 & 0.0360 & 0.0362 & 0.0298
\\
GraphDRP(GAT\_GCN) & Pearson$\uparrow$ & 0.9337 & 0.9255 & 0.7163 & 0.8428 & 0.9220 & 0.8681
\\
\citep{110bGCNforDRP} & Spearman$\uparrow$ & 0.9167 & 0.9016 & 0.5667 & 0.8639 & 0.9054 & 0.8308
\\ \midrule
\specialrule{0em}{1.5pt}{1.5pt}
\midrule
 & RMSE$\downarrow$ & 0.0236 & 0.0248 & 0.0286 & 0.0355 & \textit{0.0340} & 0.0293
\\
DeepTTA & Pearson$\uparrow$ & 0.9321 & 0.9278 & 0.7190 & 0.8487 & \textbf{0.9358} & 0.8727
\\
\citep{DeepTTC} & Spearman$\uparrow$ & 0.9134 & 0.9032 & 0.5746 & 0.8675 & \textbf{0.9181} & 0.8354
\\ \hline

& RMSE$\downarrow$ & 0.0749 & 	0.0265 & 	0.2546 & 	0.0457 & 	0.1018 & 	0.1007 &

\\

 SubCDR & Pearson$\uparrow$ & 0.9264 & 	0.9184 & 	0.1582 & 	\textbf{0.8739} & 	0.3900 & 	0.6534  &

\\

\citep{SubCDR} & Spearman$\uparrow$ & 0.8314 & 	0.8914 & 	0.1057 & 	0.8295 & 	0.3808 & 	0.6078

\\ \hline

 & RMSE$\downarrow$ & \textit{0.0217} & \textit{0.0238} & \textit{0.0282} & \textit{0.0346} & 0.0352 & \textit{0.0287}
\\
GraTransDRP & Pearson$\uparrow$ & \textit{0.9433} & \textit{0.9341} & \textit{0.7214} & 0.8559 & 0.9296 & \textit{0.8769}
\\
\citep{GraTransDRP} & Spearman$\uparrow$ & \textit{0.9277} & \textit{0.9116} & \textbf{0.5796} & \textit{0.8775} & 0.9100 & \textit{0.8413}
\\ \hline
 & RMSE$\downarrow$ & \textbf{0.0208} & \textbf{0.0229} & \textbf{0.0282} & \textbf{0.0339} & \textbf{0.0338} & \textbf{0.0279}
\\
TransEDRP(Ours) & Pearson$\uparrow$ & \textbf{0.9476} & \textbf{0.9392} & \textbf{0.7223} & \textit{0.8632} & \textit{0.9320} & \textbf{0.8809}
\\
 & Spearman$\uparrow$ & \textbf{0.9324} & \textbf{0.9170} & \textit{0.5769} & \textbf{0.8851} & 0.9128 & \textbf{0.8448}
\\ \toprule[1pt]
\end{tabular}%
}

\label{table: comparative experiment}
\end{table*}

To achieve the key goals of this paper, which are the utilization of a graph transformer with edge embedding to obtain the pharmacological and chemical features of drug molecules and the utilization of a transformer-based genomics encoder to globally extract genomic sequences, we design an overall experiment to demonstrate the accuracy advantage of TransEDRP.

In the overall experiment, several existing representative methods are selected for comparison. The overall experiments are conducted on five datasets (see Appendix Table \ref{tab:dataset}). Specifically, GDSCv1, GDSCv2, and CTRPv2 use the Hold-out method, with 80\% of data assigned to the training set, 10\% to the validation set, and 10\% to the test set. CTRPv1 and GCSI are small datasets; thus, we opt to use the K-fold method, where K is 5. Due to the limited space of the table and the small variance differences of these indicators, we have only reported the mean values of the indicators in the table.

\begin{figure}[ht!]
    \centering
    \includegraphics[width=0.65\textwidth]{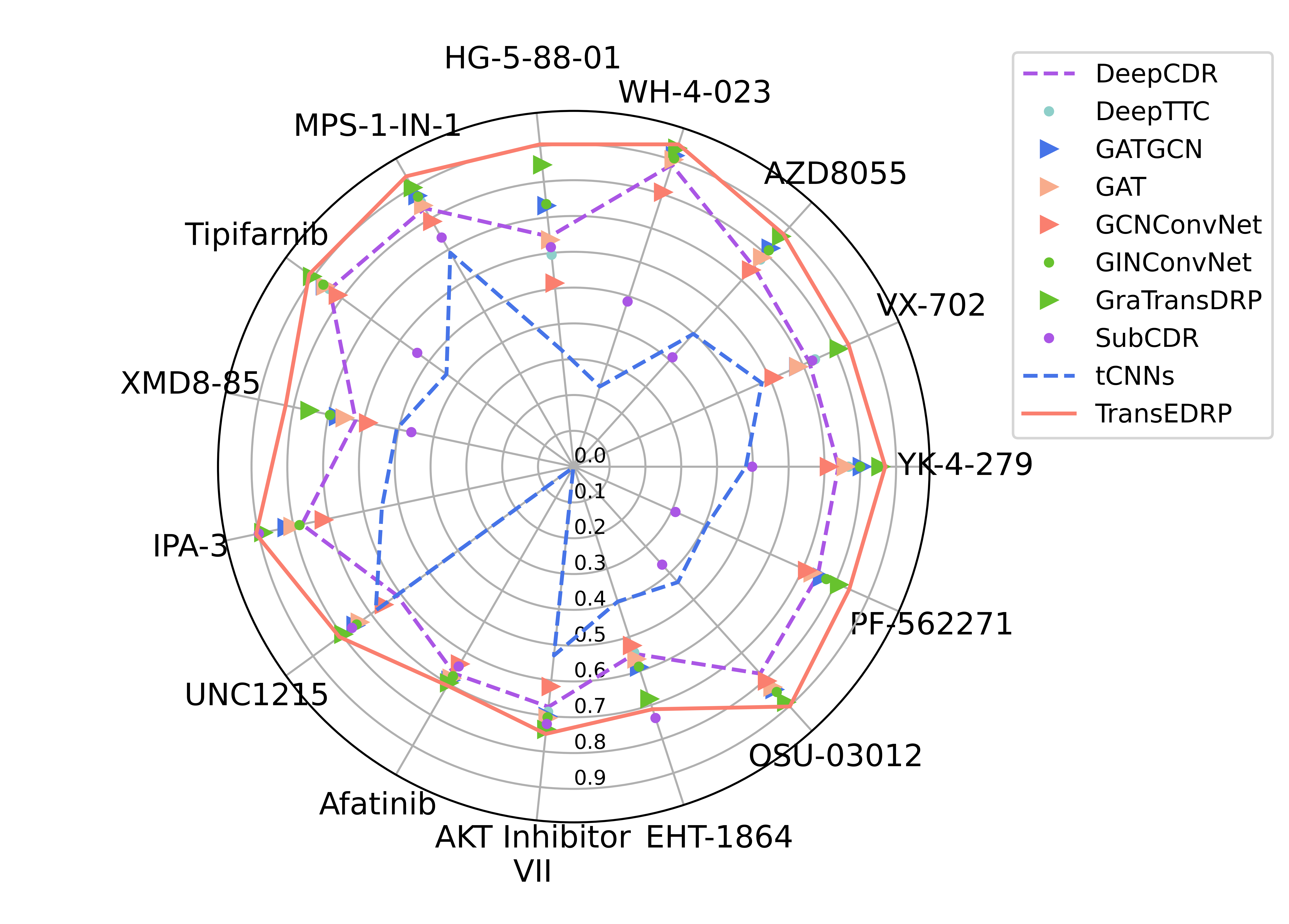}
    \caption{The figure shows the response prediction results for 15 classical drugs on GDSCv2 by mainstream methods, where $Pearson$ is chosen as the evaluation metric.
 }
    \label{fig:rader_test_show}
\end{figure}

As shown by the results in Table \ref{table: comparative experiment}, our model achieves an average improvement of \textbf{22.63\%} over the 1DCNN-based method tCNNs, \textbf{2.78\%} over the best graph-based method GraphDRP(GIN), and \textbf{1.22\%} over the best transformer-based (SOTA) method GraTransDRP. Overall, the graph-based methods are better than the non-graph-based methods due to their efficient acquisition of global information and molecular structure features. TransEDRP is the first method to embed chemical bonds as edges of graphs for representation, which makes full use of the information of drug molecules and effectively improves prediction accuracy.

In addition, the current mainstream evaluation approach calculates the macroscopic performance of mixed data for more than one drug. This macroscopic evaluation approach is more inclusive of a small number of less effective predictions and is less able to objectively assess the overall performance of the model. Therefore, we adopt a drug-by-drug type evaluation approach based on the traditional macroscopic evaluation approach, which means that the response prediction results for each drug are calculated separately and then averaged. Appendix Table \ref{tab:testshow} shows the results of the evaluation method by drug class, where \textbf{Ranking} represents the ranking loss function (the two evaluation methods do not constrain this metric). Furthermore, this paper shows in detail the prediction results of the mainstream methods on each of the 15 classical drugs. Fig. \ref{fig:rader_test_show} shows that our model is more competitive compared to the other mainstream methods on each of the 15 classical drugs (see Appendix Table \ref{tab:testshow} for details).

\subsection{Comparative experiment}
\label{sec:compare}

\begin{table*}[t!]
\centering
\caption{Comparative experiments with methods utilizing single-omics. Feasibility, stability, and safety are denoted as \textbf{F}, \textbf{St}, and \textbf{Sa}. respectively. When the type of input data is a sequence, it is abbreviated as \textbf{Seq}.}
\label{tab:comparative_single}
\setlength{\tabcolsep}{3pt}
\renewcommand\arraystretch{1.3}
\resizebox{\textwidth}{!}{%
\begin{tabular}{ccccccccccc}
\hline
\multirow{2}{*}{Methods} & \multicolumn{2}{c}{Cell} & \multicolumn{2}{c}{Drug} & \multicolumn{3}{c}{Performances on GDSCv2} & \multirow{2}{*}{F$\uparrow$} & \multirow{2}{*}{St(\textperthousand)$\downarrow$} & \multirow{2}{*}{Sa$\uparrow$} \\ \cline{2-8}
 & Input & Encoder & Input & Encoder & RMSE$\downarrow$  & PCC$\uparrow$ & Rank$\downarrow$ &  &  &  \\ \hline
tCNNs\citep{tcnns} & Seq & CNN & Fragment & CNN & 0.035  & 0.824 & 0.0326 & 0 & 57.1
 & 0 \\
DeepCDR\citep{DeepCDR} & Seq & FCN & Graph & GNN & 0.026  & 0.890 & 0.0087 & 1 & 16.8 & 0 \\
GraphDRP(GCN)\citep{GraphDRP} & Seq & CNN & Graph & GNN & 0.027  & 0.889 & 0.0085 & 1 & 12.1 & 0 \\
GraphDRP(GIN)\citep{GraphDRP} & Seq & CNN & Graph & GNN & 0.025  & 0.903 & 0.0083 & 1 & 13.3 & 0 \\
GraphDRP(GAT)\citep{GraphDRP} & Seq & CNN & Graph & GNN & 0.025  & 0.898 & 0.0080 & 1 & 11.7 & 0 \\
GraphDRP(GAT\_GCN)\citep{GraphDRP} & Seq & CNN & Graph & GNN & 0.025  & 0.901 & 0.0076 & 1 & 12.0 & 0 \\
DeepTTA\citep{DeepTTC} & Seq & FCN & Fragment & GNN & 0.024  & 0.903 & 0.0079 & 0 & \textbf{11.4} & 0 \\
SubCDR\citep{SubCDR} & Seq & Transformer & Fragment & GNN & 0.026  & 0.891 & 0.0095 & 0 & 158 & 0.5 \\
GraTransDRP\citep{GraTransDRP} & Seq & CNN & Graph & GNN & {0.023}  & {0.911} & {0.0067} & 1 & 12.0 & 0 \\
TransEDRP(Ours) & Seq & Transformer & Graph & GNN & \textbf{0.023}  & \textbf{0.917} & \textbf{0.0062} & \textbf{1} & 12.0 & \textbf{1} \\ \hline
\end{tabular}%
}
\end{table*}

\begin{table*}[t!]
\centering
\caption{Comparative experiments with methods utilizing multi-omics.}
\label{tab:comparative_multi}
\setlength{\tabcolsep}{5pt}
\renewcommand\arraystretch{1.3}

\resizebox{\textwidth}{!}{%
\begin{tabular}{cccccccccc}

\hline

\multirow{2}{*}{Methods} & \multicolumn{2}{c}{Cell} & \multicolumn{2}{c}{Drug} & \multicolumn{3}{c}{Performances on GDSCv2} & \multirow{2}{*}{F$\uparrow$} & \multirow{2}{*}{Sa$\uparrow$} \\ \cline{2-8}
 & Input & Encoder & Input & Encoder & RMSE$\downarrow$ & MAE$\downarrow$ & PCC$\uparrow$ &  &  \\ \hline
MOLI\citep{MOLI} & Seq & FCN & - & - & 1.130 & 0.875 & 0.931 & 1 & 0.5 \\
TGSA\citep{TGSA} & Graph & GNN & Graph & multi-scale GNN & 0.911 & 0.650 & 0.950 & 0 & 0.5 \\
HMM-GDAN\citep{HMM_GDAN} & Graph & GNN & Graph & multi-scale GNN & \textbf{0.869} & 0.641 & \textbf{0.952} & 0 & 0.5 \\
TransEDRP(Ours) & Seq & Transformer & Graph & single-scale GNN & {0.879} & \textbf{0.640} & {0.951} & \textbf{1} & \textbf{1} \\
\hline
\end{tabular}%
}
\end{table*}

\begin{table*}[ht!]
\caption{The table shows the performance comparison of mainstream DRP methods, where the \textbf{APP} means the average rate of processing parameters.}  
\setlength{\tabcolsep}{3pt}
\renewcommand\arraystretch{1.2}
\resizebox{\textwidth}{!}{%
\begin{tabular}{cccccc}
\toprule[1pt]
Method & Infer Time(ms) & FLOPs($10^6$) & Params(Mb) & Complexity & APP(Mb/ms) $\uparrow$  \\ \hline
tCNNS & 15 & 3.04 & 2.76 & $\mathcal{O}(9knd^2)$ & 0.179 \\
GraTransDRP & 209 & 31.34 & 7.45 &  $\mathcal{O}(3knd^{2}+(2+3d)n^{2}+3n)$  & 0.036 
\\
GraphDRP(GAT) & 4 & 5.10 & 1.29 & $\mathcal{O}(3knd^{2}+(2+2d)n^{2}+3n)$& 0.260
\\
GraphDRP(GAT\_GCN) & 20 & 3.16 & 3.57 & $\mathcal{O}(3knd^{2}+(2+d)n^{2}+3n)$ & 0.178 
\\
GraphDRP(GCN) & 4 & 2.55 & 1.29 & $\mathcal{O}(3knd^{2}+3n^{2}+3n)$ & 0.264 
\\
GraphDRP(GIN) & 4 & 2.53 & 0.83 & $\mathcal{O}(3knd^{2}+5n^{2}+3n)$& 0.177
\\
DeepCDR & 5 & 0.14 & 0.26 & $\mathcal{O}(7knd^{2}+4n^{2})$& 0.050 
\\
SubCDR & 54 & 0.99 & 3.96 & $\mathcal{O}((2k+1)nd^{2}+5nd)$& 0.073 
\\
DeepTTA & 17 & 17.56 & 4.39 & $\mathcal{O}(8knd^{2}+dn^{2})$& 0.244 
\\ \hline
TransEDRP(Ours) & 75 & 1.12 & 16.89 & $\mathcal{O}((5d+2)n^{2}+3n)$ & 0.225 
\\\toprule[1pt]
\end{tabular}%
}
\label{TabComplexity}
\end{table*}

The evaluation of DRP methods should not just focus on metrics, but also consider their performance in chirality safety and various data set scenarios, as well as their ability to handle unknown compounds effectively. We conducted comparative experiments to assess their practical effectiveness, considering feasibility, stability, safety, and model metrics. We set unified evaluation criteria for measuring their performance in these aspects.

\begin{itemize}
    \item \textbf{Feasibility}: This refers to how easily a method can screen and recommend unknown compounds in real-world applications.  If a method only considers limited training data during encoding, issues like inconsistent vector sizes may occur when new compounds or cell lines are added. According to this rule, a method that simultaneously meets the criteria of directly and concisely encoding new compounds and cell lines receives a score of 1, a score of 0.5 is assigned if only one criterion is fulfilled, and a score of 0 if neither criterion is fulfilled.


    \item \textbf{Stability}: The mean of the coefficient of variation of the three indicators on the five datasets, which measures a method's consistency across different scales, $IC_{50}$ calculation methods, and validation approaches. Methods with higher stability provide more reliable predictions for unlabeled samples. 
    

    \item \textbf{Safety}: Safety primarily considers whether a method considers the differences in drug reactions of different chiral molecules, while also considering the atomic chirality of drug molecules and the S/R configuration of chemical bonds. A score of 1 is assigned when either criterion is fulfilled, a score of 0.5 is assigned when only one criterion is fulfilled, and a score of 0 is assigned when neither criterion is fulfilled.
\end{itemize}





The comparative analysis results based on single-omics and multi-omics methods are presented in Table \ref{tab:comparative_single} and Table \ref{tab:comparative_multi}, respectively. All methods are evaluated on the GDSCv2. In the single-omics methods, tCNNs, DeepTTA, and SubCDR encode drugs using molecular fragments or frequencies. However, these methods may not be compatible with encoding new molecules, resulting in a feasibility score of 0. 

On the other hand, in the multi-omics methods, we select MOLI, TGSA, and HMM-GDAN, as shown in Table \ref{tab:comparative_multi}.  Both TGSA and HMM-GDAN abstract cell lines as graphs based on the STRING protein-protein association network, hence these methods may not be compatible with encoding new cell lines. These methods only consider the atomic chirality of drugs, resulting in a safety score of 0.5. In contrast to our approach, they introduce additional prior knowledge and employ complex preprocessing methods, making their direct applicability in practical scenarios more challenging.  It is worth noting that even though our method simply concatenates the multi-omics features of cell lines, its performance is comparable to the SOTA method HMM-GDAN.

\subsection{Ablation experiment}

In the ablation experiment, we need to prove the validity of each component of our method. In summary, we test the validity of the followings.

\begin{itemize}
    \item \textbf{V1}: The validity of the transformer encoder layers for the representation of molecular graphs;
    \item \textbf{V2}: The validity of  edge embedding in the graph representation;
    \item \textbf{V3}: The validity of multi-head attention for the extraction of cell line genomics.
\end{itemize}

In the drug branch, we introduce chemical bonds as edge embedding to participate in the message passing of node features, along with the transformer encoder layers for the representation of molecular graphs. To verify \textbf{V1} and \textbf{V2}, we design ablation experiments for the edge embedding and graph transformer encoder. 
As shown in Table \ref{tab:result}, for the \textit{Graph Transformer} and \textit{Edge Embedding}, when the other components are determined, adding these two components achieve better performance on all three metrics, which indicates that these two components can effectively mine the pharmacological features of drugs. As shown in Table \ref{tab:result}, when the other modules in TransEDRP are determined, the models that use only the graph transformer with edge embedding or only multi-head attention have a better performance in all three metrics, so \textbf{V1} and \textbf{V2} can be verified.

\begin{table}[htbp]
\caption{Ablation experiments on the GDSCv2 dataset. }

\centering
\renewcommand\arraystretch{1.2}
\setlength{\tabcolsep}{18pt}
\resizebox{\columnwidth}{!}{%
\begin{tabular}{cccccc}
\toprule[1pt]
\begin{tabular}[c]{@{}c@{}}Graph\\ Transformer\end{tabular} & \begin{tabular}[c]{@{}c@{}}Edge\\ Embedding\end{tabular} & \begin{tabular}[c]{@{}c@{}}Cell Line\\ Extract\end{tabular} & RMSE$\downarrow$ & Pearson$\uparrow$ & Spearman$\uparrow$ \\ \hline
\multicolumn{1}{c}{\multirow{4}{*}{No}} & No & Conv1d &  0.0246 & 0.9300 & 0.9097 \\
\multicolumn{1}{c}{} & Yes & Conv1d &  0.0244 & 0.9311 & 0.9105 \\
\multicolumn{1}{c}{} & No & Attention & 0.0254 & 0.9253 & 0.9037 \\
\multicolumn{1}{c}{} & Yes & Attention & 0.0252 & 0.9264 & 0.9055 \\ \hline
\multicolumn{1}{c}{\multirow{4}{*}{Yes}} & No & Conv1d & 0.0238 & 0.9348 & 0.9136 \\
\multicolumn{1}{c}{} & Yes & Conv1d & 0.0237 & 0.9353 & 0.9147 \\
\multicolumn{1}{c}{} & No & Attention &  0.0233 & 0.9377 & 0.9174 \\
\multicolumn{1}{c}{} & Yes & Attention & \textbf{0.0233} & \textbf{0.9381} & \textbf{0.9180} \\ \toprule[1pt]
\end{tabular}%
}

\label{tab:result}
\end{table}

In response to \textbf{V3}, we conduct ablation experiments on the gene sequence feature extraction methods to verify that the attention mechanism has a more complete perception field than the traditional 1D-CNN method for feature extraction. In Table \ref{tab:result}, \textit{Cell Line Extract} shows that the effectiveness of using the multi-head attention to extract cell lines genomics is only manifested when the graph transformer is used in the drug branch. Although the dual structure shows that the features of the two branches are not coupled
before fusion, the results reveal the existence of a clear interaction between them, which may imply that the extent of drug molecule representation is a determining factor in the performance of DRP models.

\subsection{Algorithm complexity}

To explore whether the prediction performance of the model causes difficulties in prediction tasks when applied to drug discovery, we conduct tests using one batch (256 samples in one batch).



Experiments are performed on a GeForce RTX 2080 Ti with inference times of 209 ms at the longest and 4 ms at the shortest.  In Table \ref{TabComplexity}, we show that the SOTA method for the drug response prediction task before our method came out is GraTransDRP, but the FLOPs overhead of GraTransDRP is too high, 28 times higher than our method, and the inference time is also 3 times longer.  If 100 million drugs are needed to predict the $IC_{50}$ for a certain cell line, the time taken is about 0.4 to 22.6 h. Drug screening relies heavily on high-precision models to provide reliable results for ranking drug responses, which can allow for this range of prediction times. In addition, inference time can be shortened by multi-GPU parallel computing. Therefore the complexity of the model and the number of parameters are acceptable in the current research approach. 

\section{Conclusion}

In this paper, we propose TransEDRP, a better model for the drug response prediction task. The TransEDRP is a dual transformer structure with edge embedding, which takes full advantage of the pharmacological and chemical properties of drug molecules, as well as extracting features of genomic sequences globally. We conduct chirality experiments and extensive overall experiments on five publicly available datasets. The results demonstrate that TransEDRP is the better model for the DRP task, which can effectively improve both the accuracy and security of drug response prediction tasks with high stability and feasibility.

\appendix


\clearpage

In this section, we will introduce the details of our model TransEDRP about several important experiments, including the case study analysis (see \ref{apd:CaseStudyAnalysis}) and the chiral security test (see \ref{apd:Chiralsecuritytest}).

\section{Pharmacy Concepts}
\label{sec:chiralConcepts}

\begin{figure}[ht!]
\centering
\includegraphics[scale=0.2]{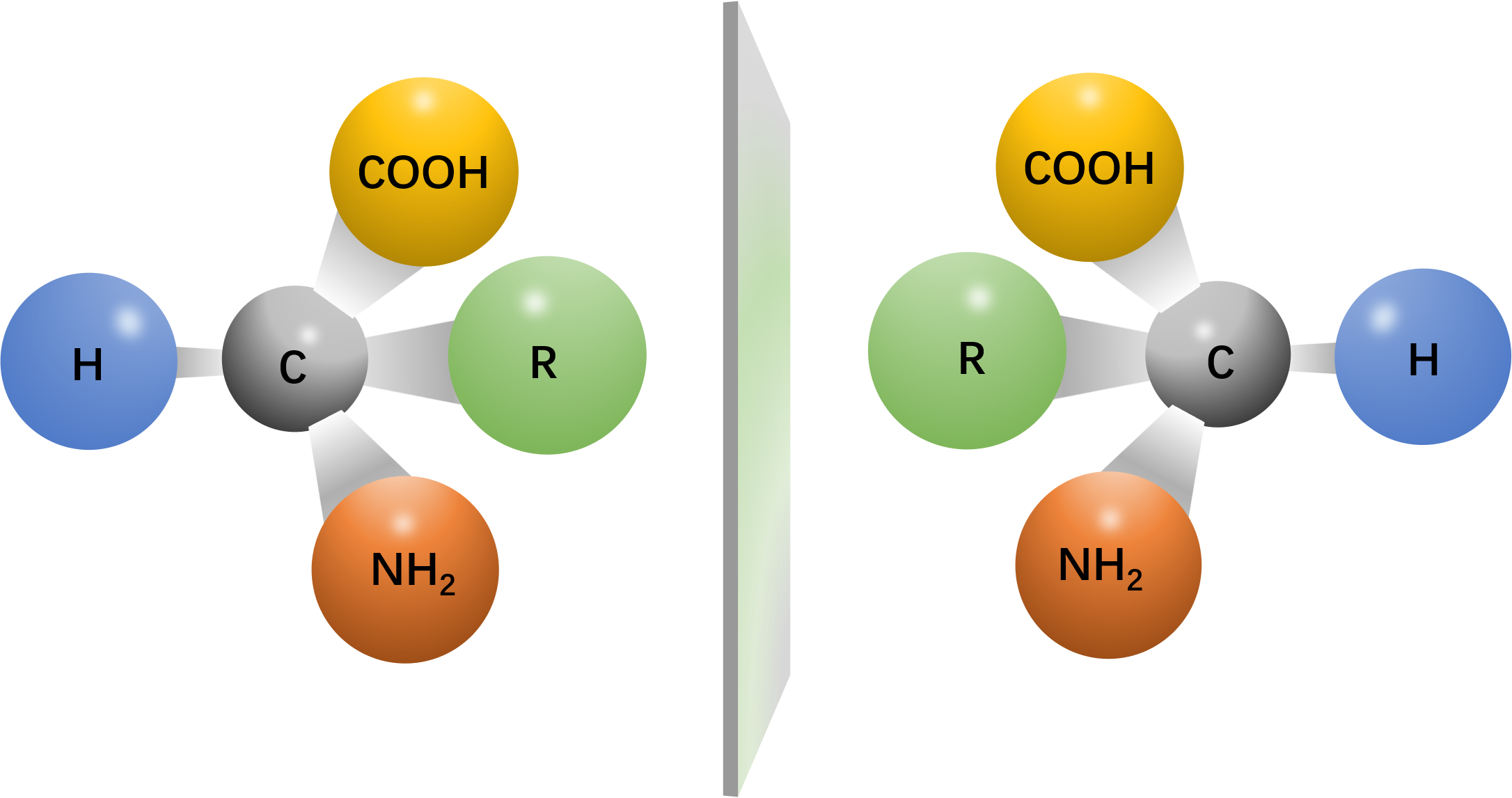}
\caption{A pair of chiral isomers, which are mirror-symmetric and cannot be overlapped.}
\label{fig-1}
\end{figure}

\noindent \textbf{Chiral Isomer.} The chiral drug is a type of drug whose molecular stereo structure and its mirror image cannot overlap with each other. A pair of drug structures that are the mirror image of each other but cannot be overlapped are called chiral isomers (enantiomers) as shown in Fig. \ref{fig-1}. Some chiral isomers of drugs and their effects are shown in Appendix Table \ref{tablechiral}, with numerous examples of how a seemingly small structural modification can make a huge difference \cite{activitycliff,ac0,ac2}. The differences in pharmacological activity between the enantiomeric isomers of chiral drugs are very different \citep{Propafenone, Flecaine, Naproxen, Chlorpheniramine}. In addition, drugs such as Vigabatrin \citep{Vigabatrin} and Methyldopa \citep{Methyldopa} have activity only in the specific isomers of levorotatory/dextrorotatory, while the remaining isomers are inactive. More importantly, the enantiomeric isomers of some drugs produce different types of activity or even opposite activity \citep{Picenadol1, Picenadol2, Penicillamine}.

\section{Experiments}\label{Experiment}

\subsection{Case Study Analysis}
\label{apd:CaseStudyAnalysis}
\begin{figure}[ht!]
    \centering
    \includegraphics[width=0.95\textwidth]{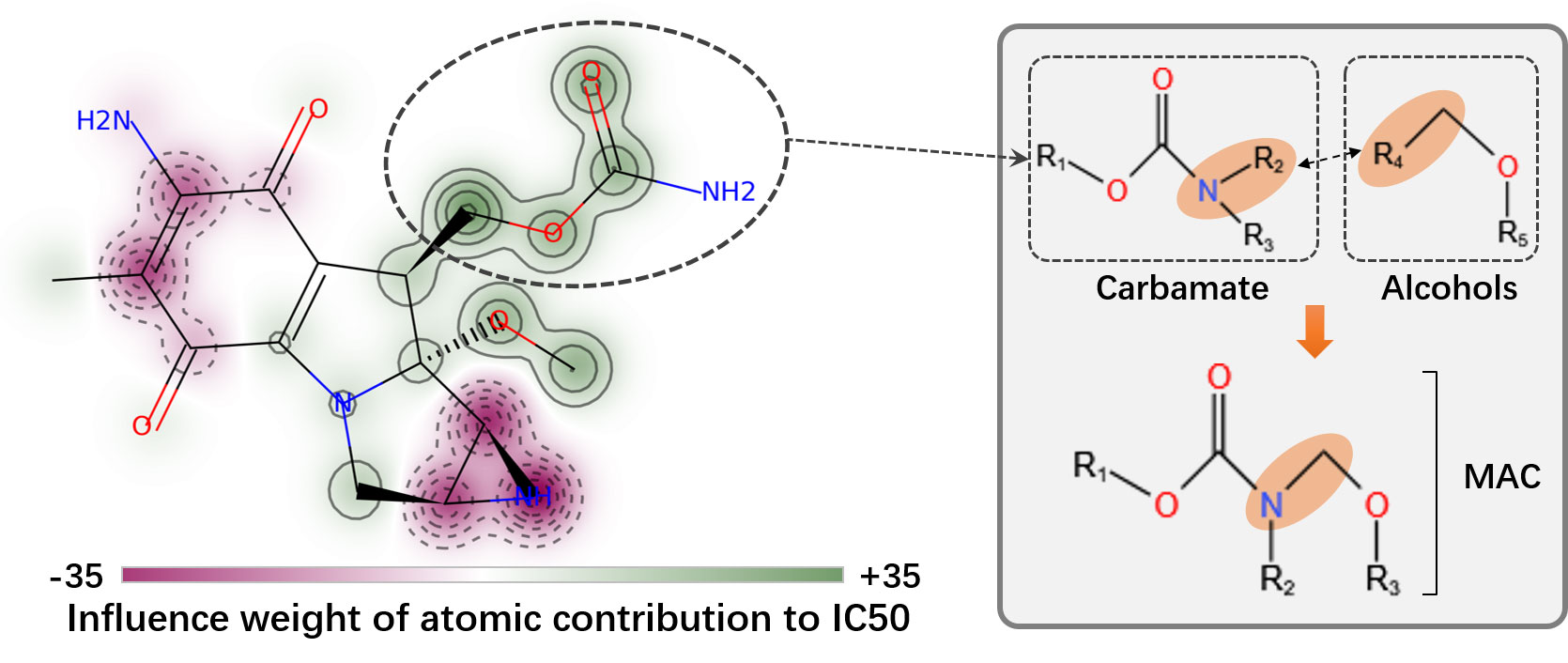}
    \caption{The figure shows 2D illustration of \textbf{Mitomycin C} (MC), where the influence weight of each atom for $IC_{50}$ is obtained by calculating the nodal feature matrix of the corresponding molecular graph from the drug branch of TransEDRP. The larger the value of the weight, the greater the contribution of that atom or group of atoms to the medicinal effect of the drug molecule. On the right side, the bioconjugation of alcohol-containing payloads to antibodies is displayed. This is achieved using a methylene alkoxy carbamate (\textbf{MAC}) self-immolative unit for alcohol conjugation. The term \textbf{Alcohols} refers to the alcohol-containing payloads.
 }
    \label{fig:mol_MC}
\end{figure}

In the overall experiment, TransEDRP shows strong accuracy and generalization, providing one SOTA DRP model in drug discovery-related studies. To verify that the drug branch of the model can effectively represent the important pharmacological properties of the drug molecular graph, we performed the case study analysis in Fig. \ref{fig:mol_MC}. This figure shows the 2D node diagram structure of the Mitomycin C molecule. We utilize TransEDRP to efficiently and automatically calculate the importance of each atom without any chemical and physical knowledge.

 

To estimate the importance of atoms, the forward propagation parameters of the Graph transformer with edge embedding module are locked and one drug molecular graph weights of this module are output according to the dimensionality of the nodes, i.e., $\mathbf{X} = \left \{ \vec{x}_1,\vec{x}_2,...,\vec{x}_N  \right \}   \in \mathbb{R}^{N\times Kd} $. Then, the feature vectors of each node $\vec{x}_i$ are summed to calculate the contribution sum of each node. Since the normalized $IC_{50}$ value ranges from 0 to 1, the closer to 0, the more effective the drug is for particular cell lines. Therefore, the mean value of the contribution of all nodes in a drug molecule is calculated, and the contribution sum of each node is subtracted from the mean value to obtain the contribution influence value of each node to the normalized $IC_{50}$ value  $I_i = \frac{  {\textstyle \sum_{i=1}^{N}} {\textstyle \sum_{j=1}^{Kd} \vec{x}_{ij}} } {N}-{\textstyle \sum_{j=1}^{Kd} \vec{x}_{ij}},i\in N$. The larger $I_i$ value indicates the greater contribution of the node to the overall potency of the drug molecule, which can be used to reveal the important atoms or atomic groups of the drug.



Mitomycin C (MC) \citep{MC3} inhibits DNA replication via Interstrand Crosslinks (ICLs) and thus prevents cancer cells from proliferating \citep{MC1, MC2}. 
As shown in Fig. \ref{fig:mol_MC}, our model clearly identifies the key functional sites within this molecule. 

The case study effectively mines the different physicochemical properties of different types of atoms of drug molecules in different structures, thus providing a reliable explanation for the different response strengths of drugs to different cell lines at the molecular level. The case study approach we performed can provide valuable interpretation and important markers for preclinical studies of drug discovery.



\subsection{Chiral security test}
\label{apd:Chiralsecuritytest}


It is well known that there are generally more than one chiral atoms in drugs. Within a drug, the combination of chiral atoms of different chiralities can lead to changes in the drug with respect to its extrinsic properties. To verify that our method can effectively reduce drug security risks by encoding the chiral features of atoms, we go on to design the chirality experiments. The previous works all do not encode the chiral features of atoms, and GraTransDRP \citep{GraTransDRP} is chosen for comparison.

\begin{figure*}[ht!]
\centering
\begin{minipage}[b]{1\linewidth}
\subfigure[\small{Chiral isomers of PFI-3.}]{
\includegraphics[width=0.49\textwidth]{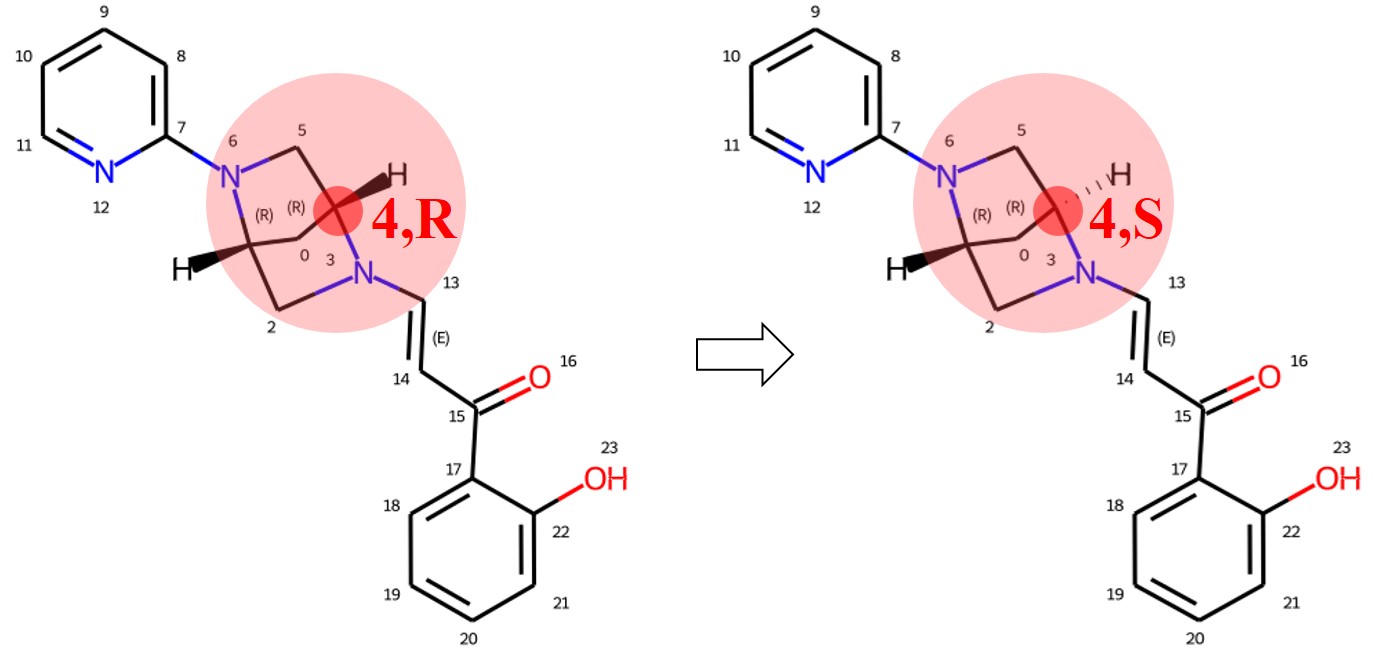}
}
\subfigure[\small{Chiral isomers of Cyclopamine.}]{
\includegraphics[width=0.49\textwidth]{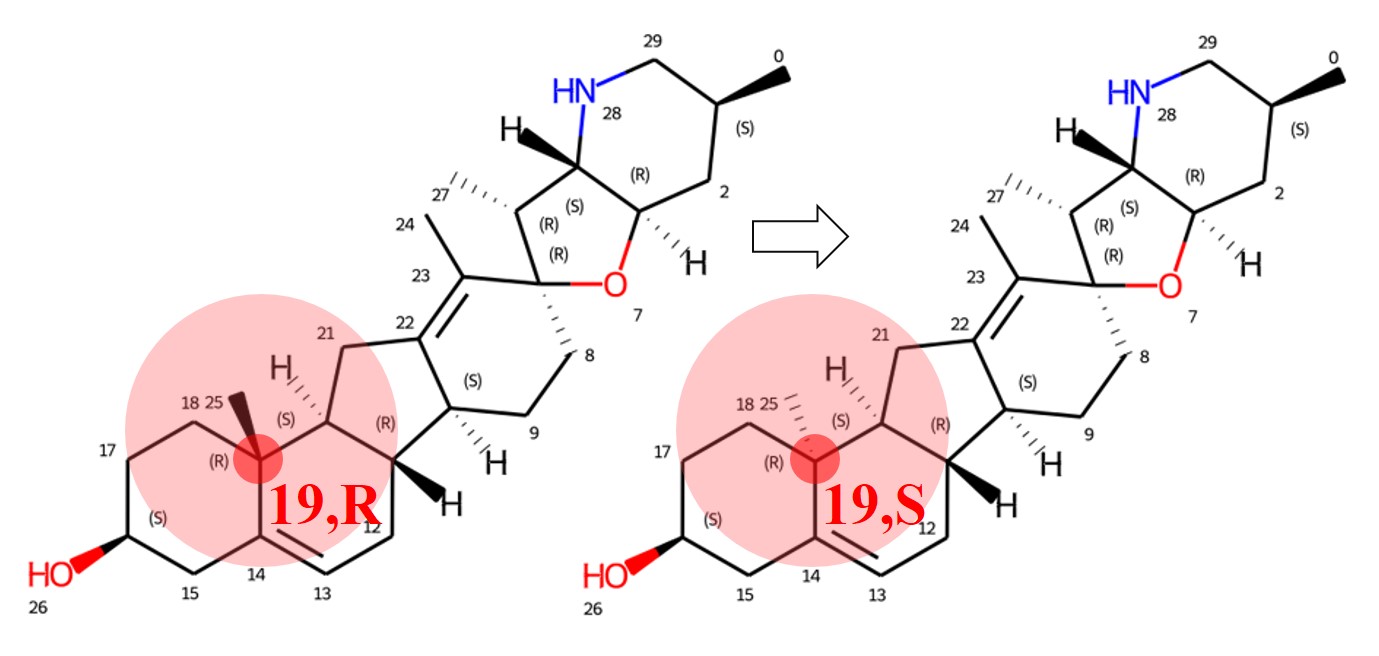}
}
\end{minipage}

\begin{minipage}[b]{1\linewidth}
\subfigure[\small{Prediction of PFI-3.}]{
\includegraphics[width=0.49\textwidth]{chirality_PFI3.jpg}
}
\subfigure[\small{Prediction of Cyclopamine.}]{
\includegraphics[width=0.49\textwidth]{chirality_cyc.jpg}
}
\end{minipage}
\caption{The prediction results of different methods on chiral isomers. The red scatterplot indicates true drug sensitivity test results,  while the dashed line indicates the prediction results.}
\label{fig:showchirality}
\end{figure*}

As shown in Fig.  \ref{fig:showchirality} (a) and (b), we select two drugs with several chiral isomers, PFI-3 and Cyclopamine; here, R/S indicates the chirality type of the carbon atoms. [$Num$, 'R'/'S'] indicates that a chiral atom is at the position of atom number $Num$, and the chiral type of the chiral atom is 'R'/'S'. To simplify the illustration and control variables to compare the perceived ability of the model for the changes of chiral information, we modified the chiral type of a single chiral atom of each drug molecule when the types of other chiral atoms are preserved consistent with the original drug. These drugs with modified chiral types and the original drugs are then fed into the model to predict the $IC_{50}$ of these drugs under a number of cell line sequences. As shown in Appendix Fig.  \ref{fig:showchirality} (c) and (d), our method can provide unique graph encoding for different chiral isomers from the same drug, which enables the model to output different prediction values based on  different inputs. However, other methods such as GraTransDRP are not sensitive to such changes, since these models do not encode the chiral features of atoms. This indicates that our method can distinguish the difference in drug sensitivity between different chiral isomers, and thus provide chirality security for drug screening.

\clearpage

\section{Tables and figures}\label{sec12}%

\begin{table*}[htbp]
\caption{A summary of the five kinds of features related to edges in a drug molecule, including the type of bond, whether it is aromatic, whether it is conjugated, whether it is in a ring, and stereo information. EDR (END-DOWN-RIGHT) and EUR (END-UP-RIGHT) are the two bond directions (for chirality).}
\renewcommand\arraystretch{1.8}

\setlength{\tabcolsep}{9pt}
\resizebox{\textwidth}{!}{%
\begin{tabular}{cccccc}
 \bottomrule 
Example & Type as Double & IsAromatic & IsConjugated & IsInRing & BondDir \\ \hline

\begin{minipage}[b]{0.15\columnwidth}
	\raisebox{-.5\height}
	{\includegraphics[width=\linewidth]{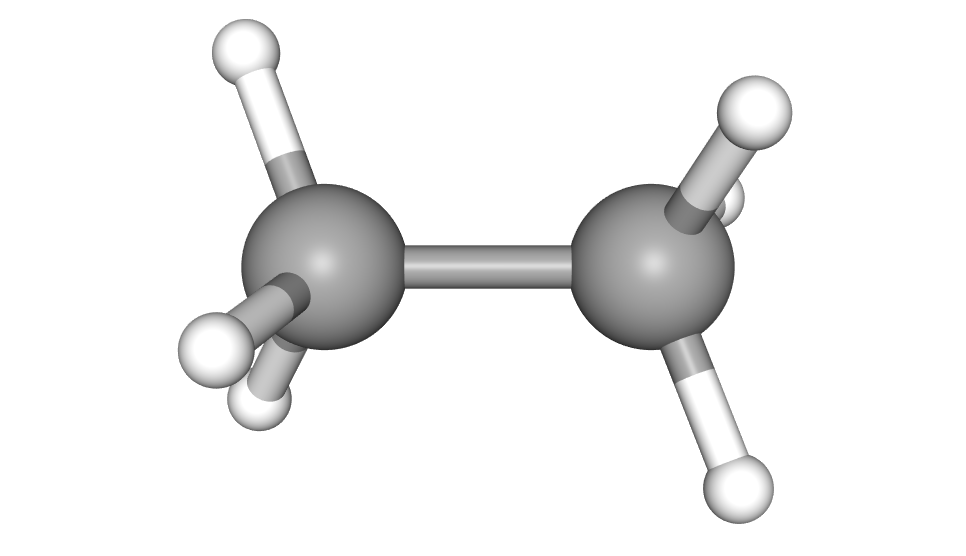}}
\end{minipage}    
	
& 1.0(Single) & False & True & False & None \\

\begin{minipage}[b]{0.15\columnwidth}
	\raisebox{-.5\height}
	{\includegraphics[width=\linewidth]{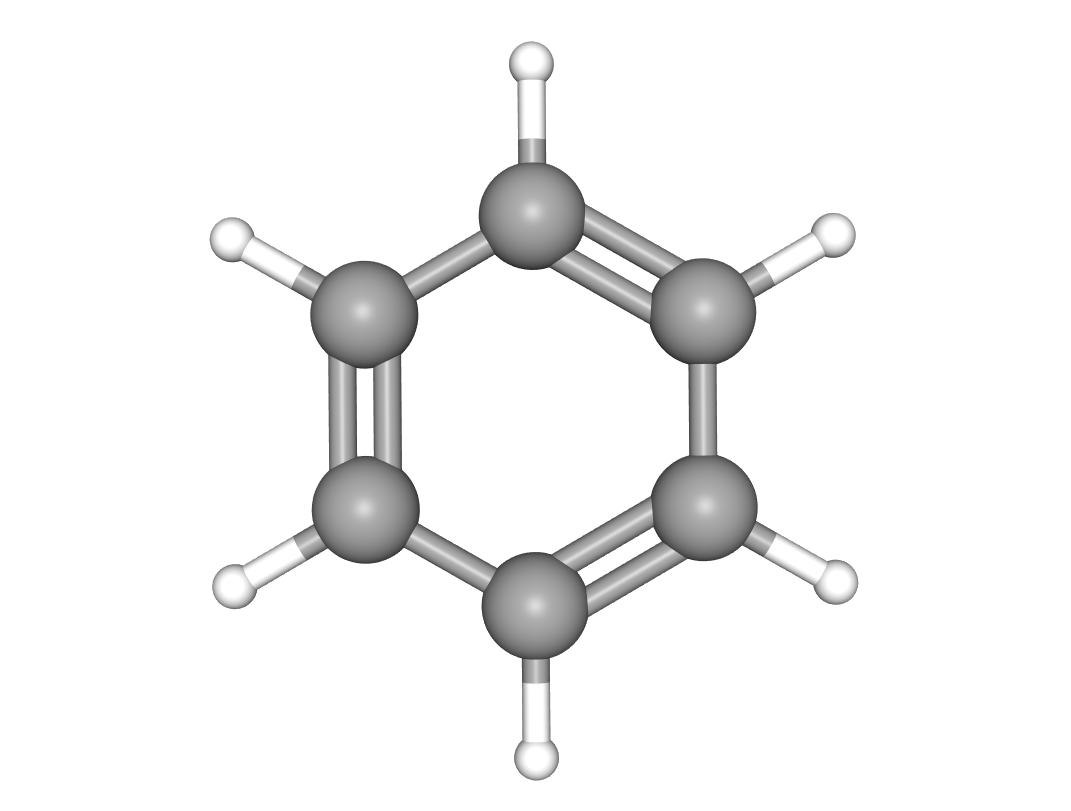}}
\end{minipage}   

& 1.5(Aromatic) & True & True & True & None \\

\begin{minipage}[b]{0.15\columnwidth}
	\raisebox{-.5\height}
	{\includegraphics[width=\linewidth]{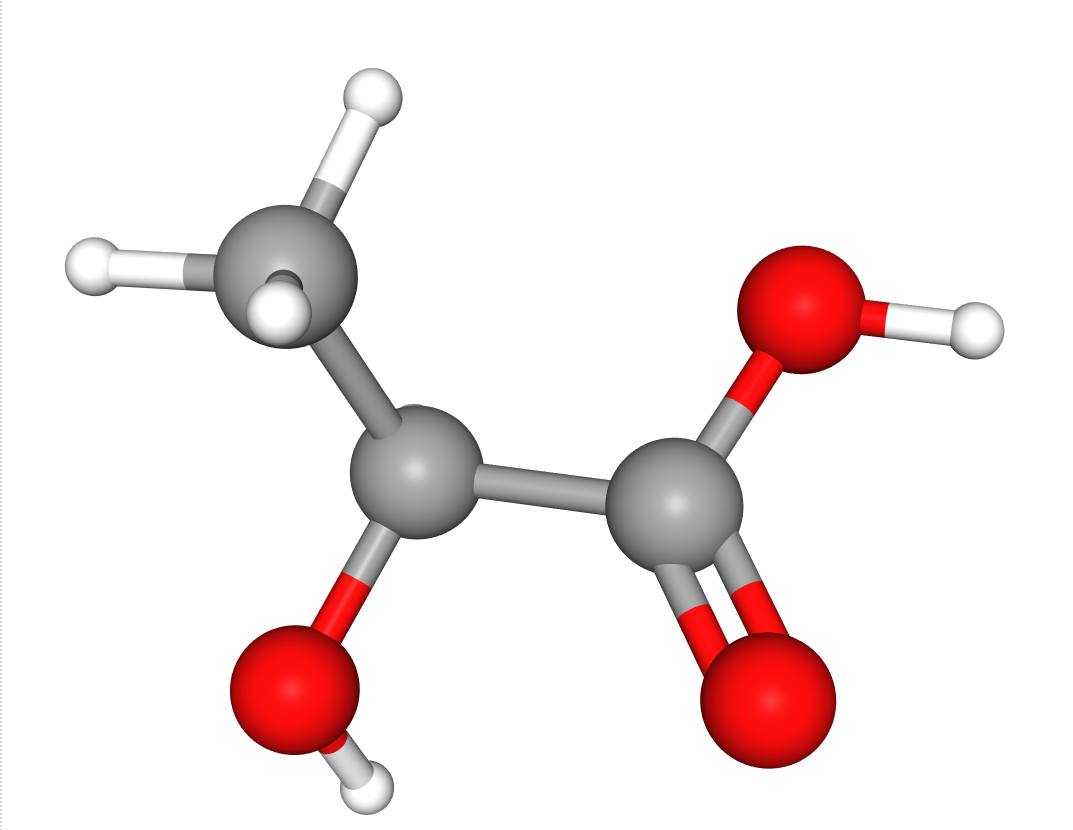}}
\end{minipage}   

& 2.0(Double) & False & True & False & EDR/EDL \\

\begin{minipage}[b]{0.15\columnwidth}
	\raisebox{-.5\height}
	{\includegraphics[width=\linewidth]{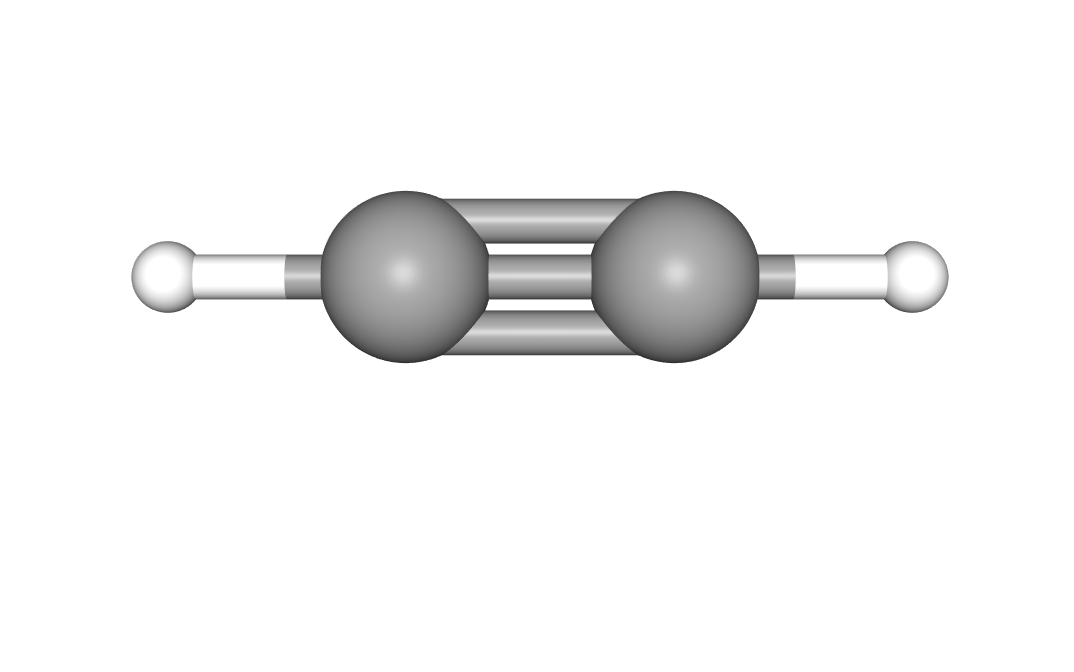}}
\end{minipage}  

& 3.0(Triple) & False & True & False & None \\  \bottomrule 
\end{tabular}%
}

\label{tab:bondE}
\end{table*}

\begin{table*}[ht!]

\caption{Examples of chiral drugs, including the effects of drugs in S/R configuration and 3D Conformer.} 
\renewcommand\arraystretch{1.2}

\setlength{\tabcolsep}{3pt}{
\resizebox{\textwidth}{!}{%
\begin{tabular}{ccccc}

\toprule[1pt]
Name & Chemical formula & S configuration & R configuration & 3D Conformer \\ \hline 
Levodopa & $C_{9}H_{11}NO_{4}$ & Parkinson's   syndrome & Granulocytopenia &  
        \begin{minipage}[b]{0.2\columnwidth}
    	\raisebox{-.5\height}
    	{\includegraphics[width=\linewidth]{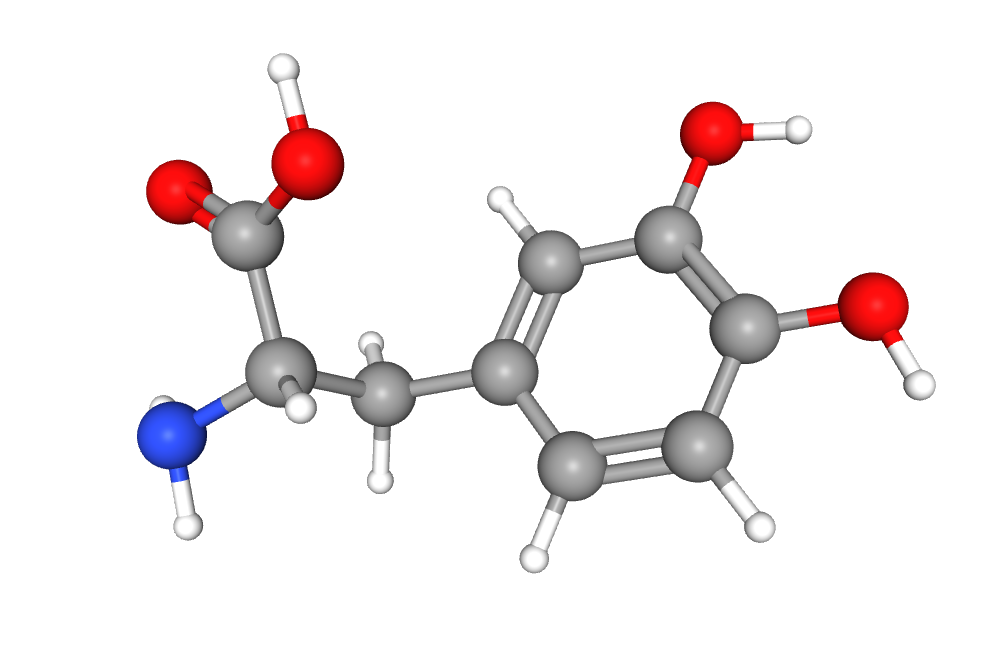}}
	\end{minipage}   \\
 
Propranolol & $C_{16}H_{21}NO_{2}$ & Beta blockers, Heart disease & Suppressing   sexual desire &  \begin{minipage}[b]{0.2\columnwidth}\raisebox{-.5\height}
    	{\includegraphics[width=\linewidth]{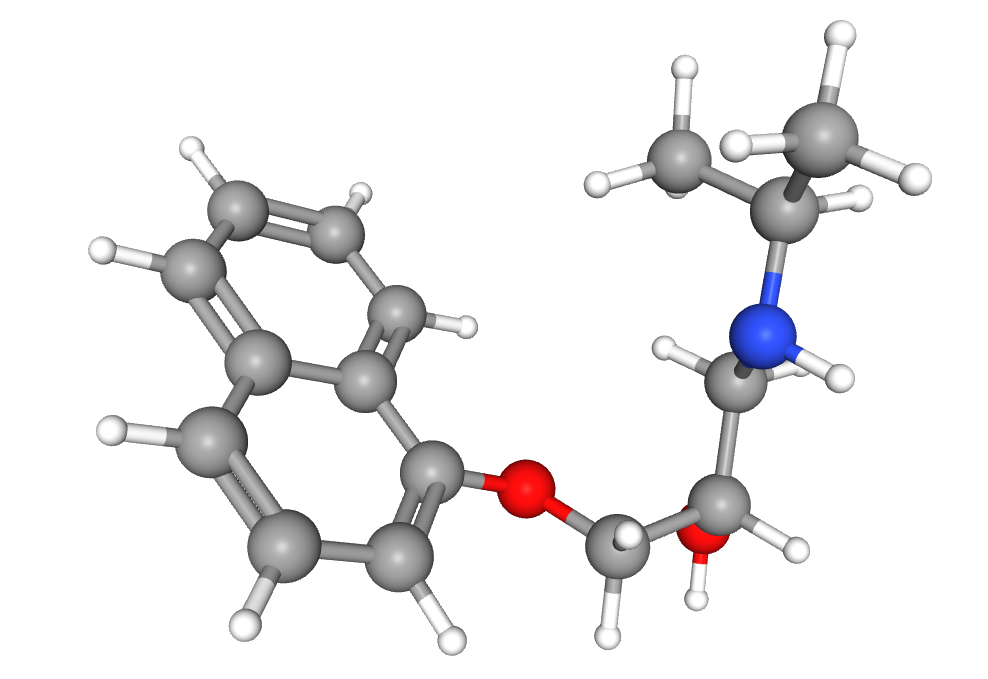}}
	\end{minipage}   \\
 
Ibuprofen & $C_{13}H_{18}O_{2}$ & \begin{tabular}[c]{@{}c@{}}Nonsteroidal,\\ Antiinflammatory Drugs\end{tabular}    & \begin{tabular}[c]{@{}c@{}}No   pharmacological,\\activity\end{tabular}  &  \begin{minipage}[b]{0.2\columnwidth}
    	\raisebox{-.5\height}
    	{\includegraphics[width=\linewidth]{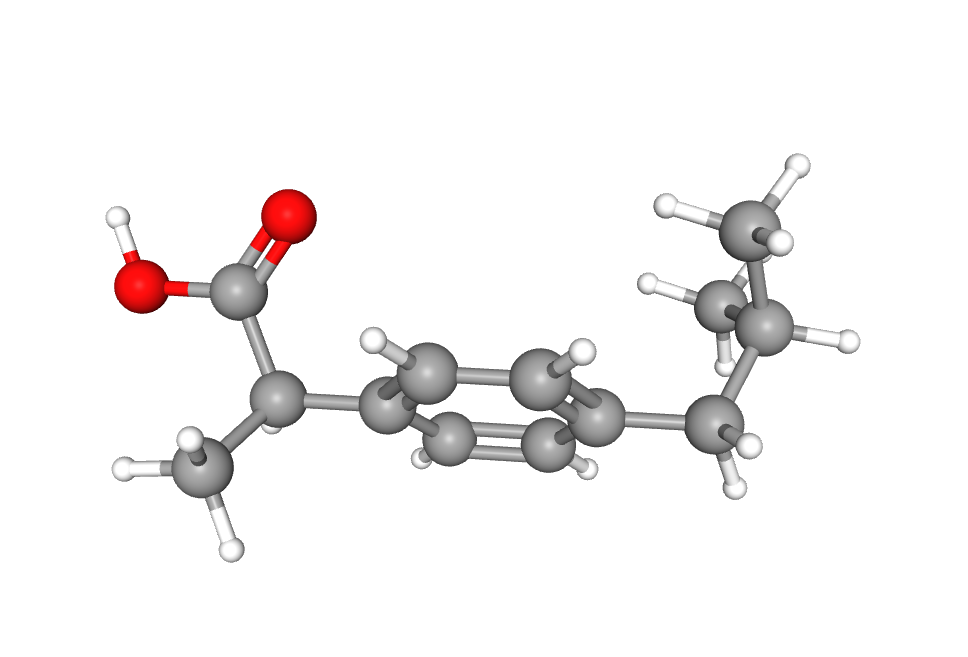}}
	\end{minipage}   \\
 
Levamisole & $C_{11}H_{12}N_{2}S$ & Anthelmintic & Vomiting &  \begin{minipage}[b]{0.2\columnwidth}
    	\raisebox{-.5\height}
    	{\includegraphics[width=\linewidth]{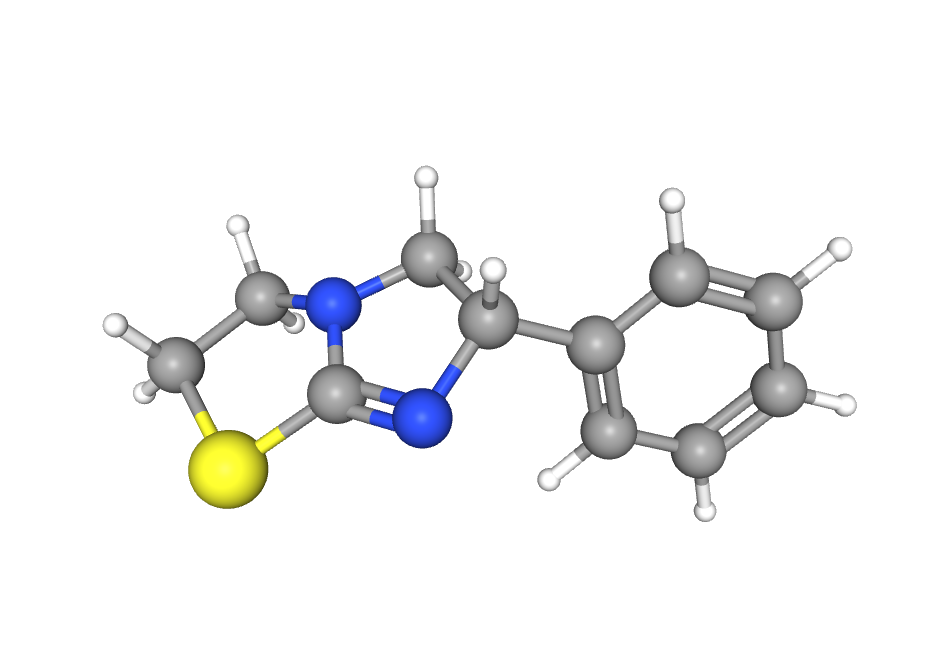}}
	\end{minipage}  \\ \toprule[1pt]
\end{tabular}%
}
}
\label{tablechiral}
\end{table*}

\begin{table*}[t!]
    \centering
    \renewcommand\arraystretch{1.5}
    \setlength{\tabcolsep}{8pt}
    \caption{Datasets for DRP tasks, where sensitivity assay means the method for calculating the $IC_{50}$.}
    \resizebox{\textwidth}{!}{%
    \begin{tabular}{cccccc}
        \hline
        Dataset     &
        GDSCv1 \citep{GDSCv1}
        &
        GDSCv2 \citep{GDSC}
        &
        CTRPv1 \citep{CTRPv1}
        &
        CTRPv2 \citep{CTRPv2_1}
        &
        GCSI \citep{GCSI_1}   \\ \hline
        Total Number     & 258196   & 152839   & 30545  & 130855   & 6178      \\ \hline
        Drug Type Number & 309      & 223      & 354    & 545      & 43        \\ \hline
        Average per drug & 835      & 655      & 86    & 240      & 143           \\ \hline
        Cell Type Number & 968      & 990      & 184    & 589      & 366           \\ \hline
        Sensitivity assay & Syto60      & CellTitreGlo      & CellTitreGlo    & CellTitreGlo      & CellTitreGlo           \\ \hline
        Model Evaluation & Hold-out & Hold-out & K-Flod & Hold-out & K-Flod \\ \hline
    \end{tabular}%
    }
    \label{tab:dataset}
\end{table*}

\begin{table}[htpb]

\caption{The table shows the average test experiments of mainstream methods on 15 classical drugs selected from GDSCv2. }
\setlength{\tabcolsep}{18pt}
\renewcommand\arraystretch{1.1}
\resizebox{\columnwidth}{!}{%
\begin{tabular}{cccccc}
\hline
Method & RMSE$\downarrow$ & Pearson$\uparrow$ & Spearman$\uparrow$ & Ranking$\downarrow$ \\ \hline
tCNNs & 0.0446 & 0.4502 & 0.4421 & 0.0326 \\
DeepCDR & 0.0218 & 0.7230 & 0.7039 & 0.0087 \\
GraphDRP(GCN) & 0.0229 & 0.6830 & 0.6646 & 0.0085 \\
GraphDRP(GIN) & 0.0200 & 0.7689 & 0.7487 & 0.0083 \\
GraphDRP(GAT) & 0.0205 & 0.7452 & 0.7265 & 0.0080 \\
GraphDRP(GATGCN) & 0.0197 & 0.7485 & 0.7675 & 0.0076 \\
DeepTTA & 0.0207 & 0.7400 & 0.7157 & 0.0079 \\
SubCDR & 0.0255 & 0.5954 & 0.5681 & 0.0095 \\
GraTransDRP & 0.0174 & 0.8246 & 0.8074 & 0.0067 \\ \hline
TransEDRP(Ours) & \textbf{0.0160} & \textbf{0.8490} & \textbf{0.8296} & \textbf{0.0062} \\ \hline
\end{tabular}%
}
\label{tab:testshow}
\end{table}

\clearpage

\bibliographystyle{elsarticle-num-names}
\bibliography{reference}

\end{document}